\documentclass[prb,floatfix,aps,twocolumn,showpacs,superscriptaddress]{revtex4-1}
\usepackage[dvips]{graphics}
\usepackage[english]{babel}
\usepackage{graphicx}
\usepackage{amssymb}
\usepackage{epstopdf}
\usepackage{latexsym}
\usepackage{amsmath}
\usepackage{MnSymbol}
\usepackage{bbold}
\usepackage{mathtools}
\usepackage{color}
\usepackage{cancel}
\usepackage{soul}

\begin{document}
\title{Vertex Representation of Hyperbolic Tensor Networks}

\author{Matej Mosko}
\affiliation{Institute of Physics, Slovak Academy of Sciences, D\'{u}bravsk\'{a} cesta 9, SK-845 11, Bratislava, Slovakia}
\author{Maria Polackova}
\affiliation{Institute of Physics, Slovak Academy of Sciences, D\'{u}bravsk\'{a} cesta 9, SK-845 11, Bratislava, Slovakia}
\affiliation{Higgs Centre for Theoretical Physics, School of Physics and Astronomy, The University of Edinburgh, Edinburgh EH9 3FD, Scotland, UK}
\author{Roman Krcmar}
\affiliation{Institute of Physics, Slovak Academy of Sciences, D\'{u}bravsk\'{a} cesta 9, SK-845 11, Bratislava, Slovakia}
\author{Andrej Gendiar}
\email{andrej.gendiar@savba.sk}
\affiliation{Institute of Physics, Slovak Academy of Sciences, D\'{u}bravsk\'{a} cesta 9, SK-845 11, Bratislava, Slovakia}

\date{\today}

\begin{abstract}
We propose a vertex representation of the tensor network (TN) for classical spin systems on hyperbolic lattices. The tensors form a network of regular $p$-sided polygons ($p>4$) with the coordination number four. The response to multi-state spin systems on the hyperbolic TN is analyzed for their entire parameter space. We show that entanglement entropy is sensitive to distinguish various hyperbolic geometries whereas other thermodynamic quantities are not. We test the numerical accuracy of vertex TNs in the phase transitions of the first, second, and infinite order at the point of maximal entanglement entropy. The hyperbolic structure of TNs induces non-critical properties in the bulk although boundary conditions significantly affect the total free energy in the thermodynamic limit. Thus developed vertex-type TN can be used for the lowest-energy quantum states on the hyperbolic lattices.
\end{abstract}


\maketitle

\section{Introduction}

The classification of phases of matter and phase transitions belongs to one of the fundamental goals in physics. Multiple analytic and numerical approaches have been developed to identify the phases of matter and classify them into universality classes. Focusing on numerical methods, we select only those that are suitable for the large number of interacting particles. In this work, we focus on improving the applicability of a numerical method that we have generalized to study models on non-Euclidean lattice geometries. Specifically, we study phase transitions induced by hyperbolic lattice geometry with constant negative Gaussian curvatures.

Among those methods, the tensor network (TN) techniques have been proven as appropriate candidates for simulating strongly correlated systems. Their popularity still increases and spreads to various fields of physics, such as condensed matter physics~\cite{TN_top_dual, Orus2}, high-energy physics~\cite{HEP1, HEP2, hp1}, quantum information~\cite{QI}, quantum machine-learning algorithms~\cite{QML, QML2}, etc. These techniques describe quantum states by a network of mutually connected tensors. Each tensor represents a vertex where a quantum particle (or a collection of particles) resides. The network of vertices imitates the interaction structure inherently incorporated into the quantum state, typically into the ground state or the lowest-lying ones. The numerical accuracy of the tensor connections is controlled by an integer parameter that we call the bond dimension (which denotes the degrees of freedom of the bond). The higher the bond dimension, the more accurately the desired quantum state is calculated.

A quantum TN state forms partially contracted tensors of a rank $(r+1)$. This is crucial since the TN geometry can be modified to create a non-Euclidean space. Let the integer $r$ represent the number of tensor indices connected with $r$ nearest-neighbor tensors (we consider a fixed coordination number). Increasing the degrees of freedom in these bond tensor indices leads to a longer range of correlations (the amount of entanglement) in a desired quantum state. The remaining tensor index is associated with the degrees of freedom of real physical particles, spin states, etc., located at the vertex.

Equivalently to the quantum state, TN can also represent the partition function of a physical system described by a classical Hamiltonian. In this work, we employ this alternative TN approach to evaluate the thermodynamic properties and entanglement entropy focusing on the properties of the underlying lattice geometry. We develop a generalized numerical algorithm based on density matrix renormalization to analyze the network structure by varying the Gaussian curvature and boundary conditions.

The TN procedure describes a process for creating large hyperbolic lattices with constant negative Gaussian curvatures on which we study multi-state spin models. We aim to investigate the influence of curved lattices on entanglement entropy (and other thermodynamic quantities).

In this study, we develop an algorithm with vertex representation on the hyperbolic lattices which has not been formulated yet. Since 2007, we have proposed weight-type representations of the hyperbolic lattices~\cite{Ueda1} that were not compatible with TN algorithms rooted in the Projected-Entangled Pair States (PEPS)~\cite{PEPS, Orus}. To accomplish the current task, we derive recurrent relations in the vertex language of the Corner Transfer Matrix Renormalization Group (CTMRG) method~\cite{Nis-ver1}. This work analyzes spin models at three types of phase transitions subject to the hyperbolic lattice geometries. In this work, we encountered irregular behavior of the entanglement entropy along with this study. It is subject to an extensive analysis that will follow soon~\cite{Vvsweight}.

The paper is organized into five sections while details are moved into appendices. In Sec.~\ref{model}, we define the two $q$-state spin models and an infinite set of hyperbolic lattices made by a regular tessellation of identical polygons (including the square lattice for benchmarking). The transformation from the weight representation into the vertex representation of the spin models is derived in Section~\ref{vertex_def}. We specify the tensors of which hyperbolic lattices are composed and set the rules (recurrent relations) necessary for building up the lattice. The specific ways of calculating magnetization and free energy are derived. Section~\ref{results} contains the analysis of the hyperbolic lattices in the vertex representation and we compare them with the weight representation we developed earlier. The results are summarized and concluded in Sec.~\ref{Concl}. We provide specific derivations in Appendices~\ref{ApA}--\ref{FEas}.

\section{Lattice Model}
\label{model}

\begin{figure}[tb]
{\centering\includegraphics[width=\linewidth]{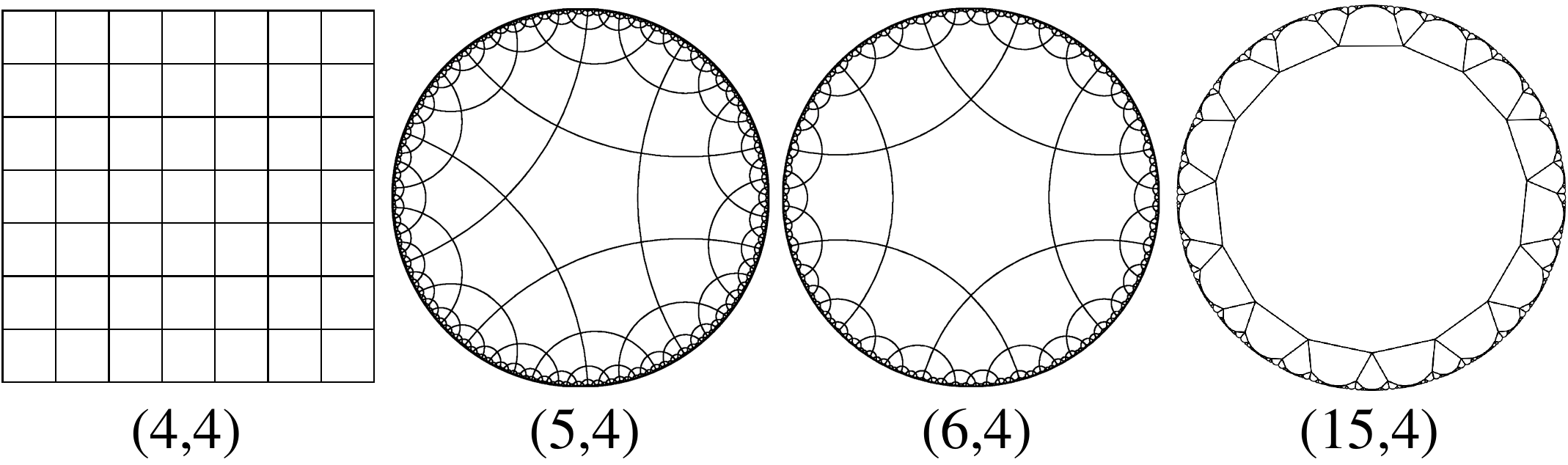}}
  \caption{The regular tessellation of the squares ($4,4$), pentagons ($5,4$), hexagons ($6,4$), and `15-gons' ($15,4$). The three hyperbolic lattices with $p\geq5$ are displayed in the Poincar\'{e} disk, i.e., the conformal representation~\cite{Poincare}.}
  \label{LatticeExamples}
\end{figure}

We study a set of lattices classified by a pair of integers ($p,r$) known as the Schl\"{a}fli symbol~\cite{Schl, Poincare}. In Fig.~\ref{LatticeExamples}, we show the Euclidean square lattice and three examples of hyperbolic lattices $(5,4)$, $(6,4)$, and $(15,4)$. We construct the lattices by the regular tessellation of congruent (uniform) polygons. The first integer $p$ describes polygons, such as, e.g., squares ($p=4$), pentagons ($p=5$), or hexagons ($p=6$), etc. The second integer $r$ is the coordination number describing the number of polygons meeting at each vertex (we consider $r=4$ fixed throughout this work; generalization to arbitrary $r$ is straightforward~\cite{weightpq}).

In this way, the infinite lattice is made by tiling the identical polygons $p$, and four polygons meet at each vertex. The vertex represents a point on the lattice where a spin is placed and interacts with the $r=4$ nearest-neighbor spins. The interactions are denoted by the lines connecting the vertices. Later on, we describe a procedure to produce these hyperbolic lattices, $(p\geq4,4)$.

The vertices contain $q$-state spins represented by variables $\sigma_{i}=0,1,\dots,q-1$ where $q=2,3,\dots$. The subscript $i$ enumerates the position of the spin variable at the vertex.

We consider two spin models with the nearest-neighboring spins: The $q$-state clock model with the planar spin-spin interaction 
\begin{equation}
    {\cal S}_{\sigma_{i}\sigma_{j}}=\cos\left[\frac{2\pi}{q}(\sigma_{i}-\sigma_{j})\right]
\end{equation}
and the $q$-state Potts model with
\begin{equation}
    {\cal S}_{\sigma_{i}\sigma_{j}}=\delta_{\sigma_{i} \sigma_{j}},
\end{equation}
where $\delta_{\sigma_{i} \sigma_{j}}$ is the Kronecker delta. Then, the two-spin local Hamiltonian between the adjacent (nearest-neighbor) spins $\sigma_{i}$ and $\sigma_{j}$ has the form
\begin{equation}
    H_{ij} = -J  {\cal S}_{\sigma_{i}\sigma_{j}}
    - \frac{h}{2}\left({\cal S}_{\sigma_{i} \vartheta} + {\cal S}_{\sigma_{j} \vartheta} \right) - b\, {\cal S}_{\sigma_{i} \vartheta}.
\label{LocHam}
\end{equation}
The uniform spin interaction $J$ acts between the nearest-neighboring spins $\sigma_{i}\sigma_{j}$. We assume two independent external magnetic fields $h$ and $b$ that we specify below. The single-state integer $\vartheta$ takes one of the $\sigma$ values being a reference-spin direction used when evaluating magnetization (we fix it to be $\vartheta=0$ in this work). The $2$-state clock and the $2$-state Potts models coincide with the Ising model after rescaling $J \to \frac{J}{2}$.

The full Hamiltonian ${\cal H}_p$ is constructed by connecting the two-spin local Hamiltonians $H_{ij}$ in Eq.~\eqref{LocHam} to comply the regular $p$-polygonal tessellation of the entire lattice $(p,4)$, i.e.,
\begin{equation}
    {\cal H}_p[\sigma]  = -J\hspace*{-0.2cm} \sum\limits_{\{\{i,j\}\}_p} \hspace*{-0.2cm}{\cal S}_{\sigma_{i}\sigma_{j}}
    - h\sum\limits_{\{i\}_p} {\cal S}_{\sigma_{i} \vartheta} - b\sum\limits_{\{i\}_p^b} {\cal S}_{\sigma_{i} \vartheta}\,.
    \label{H}
\end{equation}
The configuration sum $\{\{i, j\}\}_p$ is appropriately taken over the nearest-neighboring spins for each lattice geometry $(p,4)$. We denote the second summing notations $\{i\}_p$ for all spins in the respective lattice geometry, where the constant magnetic field $h$ is imposed on every spin (vertex), whereas $\{i\}_p^b$ denotes the sum over the spins located on the lattice boundary only, where the magnetic field $b$ is imposed. The magnetic field $h$ also acts on the boundary, cf. Eq.~\eqref{LocHam}.

We introduce a simplified notation in which $[\sigma]$ means grouping a set of all ${\cal N}$ spins that are used to construct the entire lattice, i.e., ${\cal H}_p[\sigma]\equiv{\cal H}_p(\sigma_1\sigma_2\cdots\sigma_{\cal N})$. We consider the thermodynamic limit (${\cal N} \rightarrow \infty$) since we intend to study the phase transitions after the symmetry is spontaneously broken.

To evaluate the partition function, we first define the local Boltzmann weight ${\cal W}_{\sigma_{i}\sigma_{j}}$ between two adjacent spins
\begin{equation}
    {\cal W}_{\sigma_{i}\sigma_{j}} = \exp{\left(- \frac{H_{ij}}{k_{\rm B}T }\right)},
\end{equation}
where $k_{\rm B}$ and $T$ are the Boltzmann constant and thermodynamic temperature, respectively. Then, the partition function of the total Hamiltonian reads
\begin{equation}
    {\cal Z}_p^{({\cal W})} = \sum\limits_{[\sigma]} e^{-{{\cal H}_p{[\sigma]}/{k_{\rm B} T}}} = \sum\limits_{[\sigma]} \prod \limits_{\{\{i,j\}\}_p} {\cal W}_{\sigma_{i} \sigma_{j}}
    \label{PF1}
\end{equation}
where the sum is taken over all spin configurations $[\sigma]$ between the nearest-neighboring spins.

\section{Vertex representation}
\label{vertex_def}

There are two equivalent representations in expressing the partition function. The first representation uses the Boltzmann weights to build up the entire lattice by taking the product of ${\cal W}$ and the sum runs over all $q^N$ spin configurations $[\sigma]$, as defined above in Eq.~\eqref{PF1}. We name this the {\it weight} representation. The second representation, namely the {\it vertex} representation, expresses the partition function as a product of rank-$r$ tensors ${\cal V}$ (defined below) where the sum runs over all bond configurations (not the spin ones). Then, the partition function is defined as
\begin{equation}
    {\cal Z}_p^{({\cal V})} = \sum\limits_{[a,b,c,d]} \prod\limits_{i} {\cal V}_{a_i,b_i,c_i,d_i} = {\rm Tr}\, \prod {\cal V}_{abcd}.
\end{equation}
In what follows, we formulate the vertex representation of the partition function to show that the equivalence ${\cal Z}_p^{({\cal W})} = {\cal Z}_p^{({\cal V})}$ holds. We thus relate the vertices ${\cal V}$ with the weights ${\cal W}$. 

Since we restrict our study to the case of $r=4$, the rank-4 tensor ${\cal V}_{abcd}$ consists of four bond variables $a,b,c,d=0,1,\dots,q-1$. Generalization to an arbitrary coordination number $r$ is easy and leads to vertices as rank-$r$ tensors. We reserve the Latin letters $a,b,\dots$ to denote the bond variables, whereas the indexed Greek letter $\sigma_i$ refers to the spin variable used in the weight representation.

\begin{figure}[tb]
{\centering\includegraphics[width=0.5\linewidth]{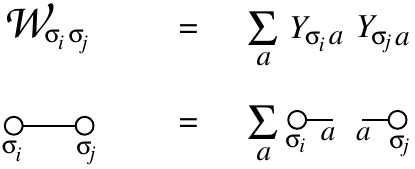}}
\caption{Graphical visualization of the Boltzmann-weight decomposition according to Eq.~\eqref{WB} used.}
\label{Wsisj}
\end{figure}

We begin by decomposing the two-spin Boltzmann weight ${\cal W}$ into the product of two identical matrices $Y$ either one having one spin index $\sigma$ and one bond index $a$, as depicted in Fig.~\ref{Wsisj}. The Boltzmann weight ${\cal W}_{\sigma_i \sigma_j}$ in the weight representation can be diagonalized since ${\cal W}$ is a real symmetric $q\times q$ matrix
\begin{equation}
    \begin{split}
   {\cal W}_{\sigma_{i}\sigma_{j}} &= \sum\limits_{a,b=0}^{q-1} U^{~}_{\sigma_i\,a}\,D^{~}_{ab}\, U^\dagger_{b\,\sigma_j} = \sum\limits_{a,b=0}^{q-1} U^{~}_{\sigma_i\,a}\,\lambda_a^{~}\delta^{~}_{ab}\, U^\dagger_{b\,\sigma_j}\\
    &= \sum\limits_{a=0}^{q-1} \left(U^{~}_{\sigma_i\,a}\sqrt{\lambda_a^{~}}\right)\left(U^{~}_{\sigma_j\,a}\sqrt{\lambda_a^{~}}\right)
    = \sum\limits_{a=0}^{q-1} Y^{~}_{\sigma_i\,a}Y^{~}_{\sigma_j\,a}.
    \label{WB} 
    \end{split}
\end{equation}
Provided that coupling is ferromagnetic, i.e., $J>0$, the diagonal matrix $D_{ab} = \lambda_a\delta_{ab}$ contains only non-negative eigenvalues $\lambda_a \geq 0$. 
Thus, we have decomposed ${\cal W}_{\sigma_i \sigma_j}$ onto two identical matrices (rank-2 tensors) $Y_{\sigma\,a}= U^{~}_{\sigma\,a}\sqrt{\lambda_a^{~}}$  with two distinct variables: the spin variable $\sigma$ and the bond variable $a$ to be used for constructing the entire TN.

Equation~\eqref{WB} describes a symmetric decomposition in the vertex representation~\cite{sqrt(Q)}, where we schematically split the Boltzmann weight  ${\cal W} = Y\cdot Y$ into the product of two identical matrices $Y$. The  decomposition holds for symmetric Hamiltonians with non-negative eigenvalues of  ${\cal W}$. If comparing the decomposition in Eq.~\eqref{WB} with that in Ref.~\onlinecite{sqrt(Q)} where ${\cal W} = {\sqrt{\cal W}}\cdot {\sqrt{\cal W}}$, they both become equivalent if $\sqrt{{\cal W}} = Y$ for boundary magnetic field $b=0$.

We intentionally impose $b\neq0$ on the boundary spins only, see Eq.~\eqref{LocHam}, to induce a faster spontaneous symmetry breaking. This leads to the asymmetric Hamiltonian and there are two ways of the decomposition ${\cal W}=Y\cdot Y'$. We use either the eigenvalue decomposition with $Y_{\sigma a}=U_{\sigma a}|d_a|$ and $Y'_{\sigma a}=\text{sgn}(d_a)Y_{\sigma a}$ or, equivalently, we can use the singular value decomposition with $Y_{\sigma a}=U_{\sigma a}\sqrt{d_a}$ and $Y'_{\sigma a}=V_{\sigma a}\sqrt{d_a}$ where all singular values are non-negative $d\geq0$ after decomposing ${\cal W}_{\sigma\sigma'}=\sum_x U_{\sigma x} d_x V_{\sigma' x}$.

\begin{figure}[tb]
{\centering\includegraphics[width=0.7\linewidth]{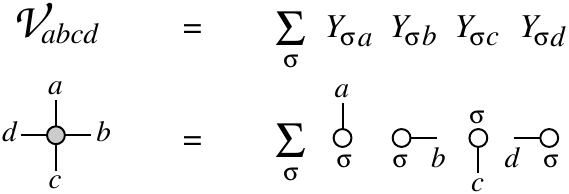}}
\caption{Visualization of the vertex construction of ${\cal V}_{abcd}$ as specified in Eq.~\eqref{Vabcd}.} 
  \label{V_abcd}
\end{figure}

\begin{figure}[!b]
{\centering\includegraphics[width=0.7\linewidth]{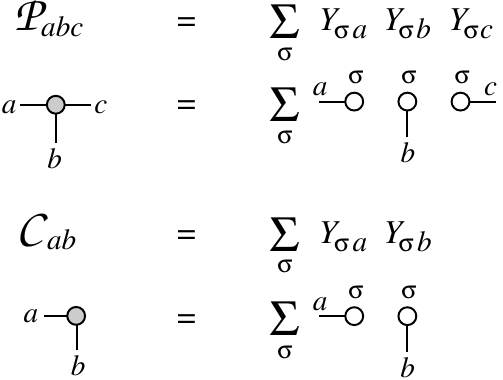}}
\caption{The initialization of the transfer tensor ${\cal P}_{abc}$ and the corner transfer matrix ${\cal C}_{ab}$.}
  \label{PabcCab}
\end{figure}

We can now define the basic vertex tensor
\begin{equation}
    {\cal V}_{abcd} = \sum\limits_{\sigma}Y_{\sigma a} Y_{\sigma b} Y_{\sigma c} Y_{\sigma d}
    \label{Vabcd}
\end{equation}
by summing up the $q$-state spin $\sigma=0,\dots,q-1$ (see the graphical visualization in Fig.~\ref{V_abcd}). The rank-$r$ tensor ${\cal V}$ is the desired {\it vertex} required in forming the vertex representation of TN. The vertex ${\cal V}$ coincides with the coordination number of the lattice (here, $r=4$). Building up a lattice with an arbitrary constant coordination number $r$ requires preparing a rank-$r$ tensor by summing the product of $r$ matrices $Y$ in Eq.~\eqref{Vabcd}. Hence, the weight and vertex representations are equivalent in the sense of ${\cal Z}_p^{({\cal W})} = {\cal Z}_p^{({\cal V})}$ which are interconnected via Eqs.~\eqref{WB} and~\eqref{Vabcd}. As specified below, we define two types of tensors: the rank-3 transfer tensor  ${\cal P}_{abc} = \sum\limits_{\sigma}Y_{\sigma a} Y_{\sigma b} Y_{\sigma c} $ and the corner transfer matrix ${\cal C}_{ab} = \sum\limits_{\sigma}Y_{\sigma a} Y_{\sigma b} $, as in Fig.~\ref{PabcCab} in analogy to the rank-4 vertex tensor ${\cal V}_{abcd}$. 

The construction of the ($p,4$) lattices follows an algorithm with the iterative scheme. Let the integer $j=1,2,3,\dots$ enumerate the iteration steps. To start, we prepare two initial tensors at the first iteration $j=1$: the rank-3 transfer tensor ${[{\cal P}_{j=1}]}_{abc}\equiv {\cal P}_{abc}$ and the rank-2 corner transfer matrix ${[{\cal C}_{j=1}]}_{ab} \equiv {\cal C}_{ab}$ which are placed on the lattice boundary and define boundary conditions. We reserve the integer $k$ to describe the final iteration step in the sequence $j=1,2,3,\dots,k$.

\subsection{Recurrent scheme}

Building up the hyperbolic lattice requires deriving {\it recurrent relations} which describe the incremental expansion of the ($p,4$) lattices gradually using $j$. To accomplish this, we employ the CTMRG method as an iterative algorithm that gradually expands lattice size beginning from the boundary towards the lattice center. At each iteration step $j$, the CTMRG procedure consists of two parts: \textit{extension} and \textit{renormalization}. The \textit{extension} forms the tensors ${[\tilde{\cal P}_{j+1}]}_{abc}$ and ${[\tilde{\cal C}_{j+1}]}_{ab}$ that are built using ${[{\cal P}_j]}_{abc}$ and ${[{\cal C}_j]}_{ab}$ from the previous iteration step. The \textit {renormalization} reduces the exponentially growing Hilbert dimension in the indices $a,b,c$, i.e., it maps the tensors onto the lower-dimensional tensors ${[\tilde{\cal P}_{j+1}]}_{abc} \to {[{\cal P}_{j+1}]}_{abc}$ and ${[\tilde{\cal C}_{j+1}]}_{ab} \to {[{\cal C}_{j+1}]}_{ab}$. This step is performed by the leading eigenvectors of the reduced density matrix $\rho'_j$ while neglecting states of the least probable configurations.

In what follows, we simplify the tensor subscript notation of ${[{\cal P}_j]}_{abc}$ and ${[{\cal C}_j]}_{ab}$ by omitting the bond indices $a,b,c$ and leaving the iteration steps $j$ only. With this simplification, we focus on the tensor extension process only. The detailed use of the bond indices in the recurrent expressions is given in Appendix~\ref{ApA}.

For instructive purposes, we begin with the recurrent scheme for the known square lattice $(4,4)$ followed by the hyperbolic pentagonal lattice $(5,4)$, see Fig.~\ref{LatticeExamples}. Afterward, we generalize the recurrent relations to the $(p\geq6,4)$ lattices.

\begin{figure}[tb]
{\centering\includegraphics[width=\linewidth]{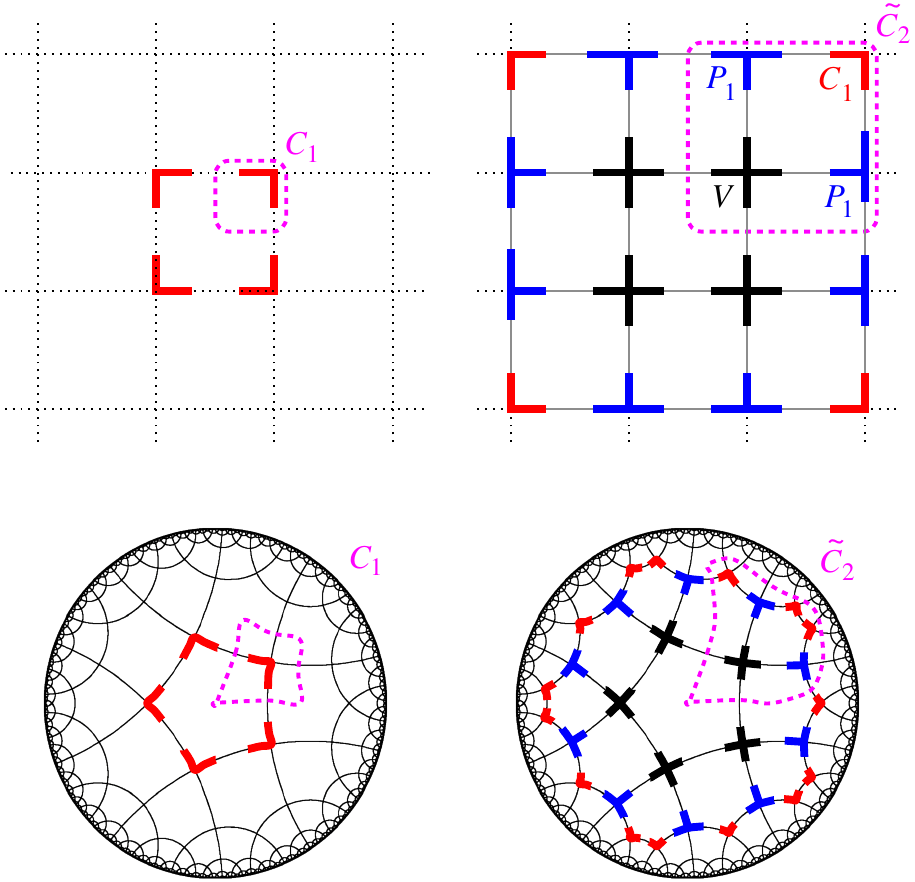}}
  \caption{Expansion process of the corner transfer tensors ${\cal C}_{j=1} \to \tilde{\cal C}_{j=2}$ on the square $(4,4)$ and hyperbolic $(5,4)$ lattices.}
  \label{C1-C2}
\end{figure}
\begin{figure}[tb]
{\centering\includegraphics[width=\linewidth]{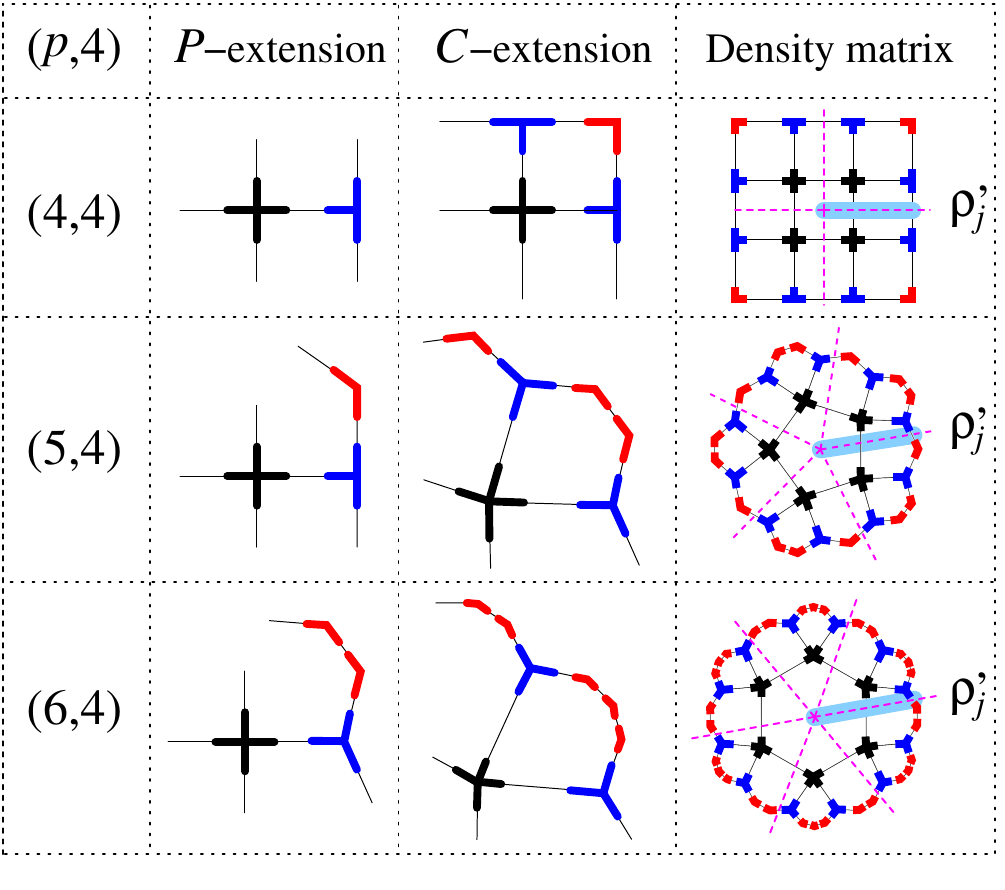}}
  \caption{A schematic visualization of the tensor extensions ${\cal P}_{j} \to \tilde{\cal P}_{j+1}$ and ${\cal C}_{j} \to \tilde{\cal C}_{j+1}$ in agreement with Eq.~\eqref{extp4short} for the lattices $(4,4)$, $(5,4)$, and $(6,4)$. The TN is made of tensor tessellations of the rank-4 $\left[{\cal V}\right]_{abcd}$ (in black), the rank-3 tensor $\left[{\cal P}_{j}\right]_{abc}$ (in blue), and the rank-2 tensor $\left[{\cal C}_{j}\right]_{ab}$ (in red). The reduced density matrix is visualized as a cut (in light blue) in the tensor network formed by the $p$ extended corner transfer matrices $\tilde{\cal C}_{j+1}$, as in Eq.~\eqref{RDM}.} 
  \label{extension}
\end{figure}
Figure~\ref{C1-C2} depicts the extension of the Euclidean square and the pentagonal lattices. The thick black crosses represent the vertex-type tensors ${\cal V}$, the thick red corners represent the two-index corner transfer tensors ${\cal C}$, and the thick blue T-shaped symbols represent the three-index transfer tensor ${\cal P}$. The ranks of the tensors ${\cal V}$, ${\cal C}$, and ${\cal P}$ remain identical for all of them. The lattice geometries ($4,4$) and ($5,4$) are illustrated by thin black lines forming congruent squares and pentagons, respectively. The spins $\sigma$ sit in the vertices and the thin lines represent the nearest-neighbor interactions connecting the vertices.

If we compare the original extension process of the square $(4,4)$ lattice~\cite{Nis-ver1, Nis-ver2, Nis-ver3} with the current notation (as visualized in Fig.~\ref{C1-C2}), we can express the recurrence relations in a simplified form $\tilde{\cal C}_{j+1}={\cal V}{\cal P}_{j}^{2} {\cal C}_{j}^{~}$ and $\tilde{\cal P}_{j+1}={\cal V} {\cal P}_{j}^{~}$ on the square lattice. Using this analogy, the expansion of the hyperbolic pentagonal $(5,4)$ lattice obeys the rules $\tilde{\cal C}_{j+1} = {\cal V}{\cal P}_{j}^{2} {\cal C}_{j}^{3}$ and $\tilde{\cal P}_{j+1}= {\cal V}{\cal P}_{j}^{~} {\cal C}_{j}^{~}$, cf.~Fig.~\ref{C1-C2}. 

Having considered the geometric structure of all hyperbolic lattices $(p,4)$, including the Euclidean $(4,4)$, we can generalize the complete class of the lattices for $p=4,5,6,\dots$ to derive the recurrence relations
\begin{equation}
\begin{split}
\tilde{\cal C}_{j+1}&={\cal V}{\cal P}_{j}^{2} {\cal C}_{j}^{2p-7}\,,\\
\tilde{\cal P}_{j+1}&={\cal V} {\cal P}_{j}^{~} {\cal C}_{j}^{p-4}\,.
\end{split}
\label{extp4short}
\end{equation}
These recurrence relations of TN are expressed concisely in Eq.~\eqref{extp4short}. They represent tensor contractions, as specified in Appendix A. (There are minor differences between the vertex recurrence relations in Eq.~\eqref{extp4short} and the weight-type ones~\cite{weightpq}.)

Now we describe how the extended tensors of rank-4 $\tilde{\cal C}_{j+1}$ and rank-5 $\tilde{\cal P}_{j+1}$ are transformed back onto the original rank-2 ${\cal C}_{j+1}$ and rank-3 ${\cal P}_{j+1}$ tensors by renormalization processes. We do this using the isometry $U_{j+1}$ (originating in the reduced density matrix) which is a unitary projection $U_{j+1}^\dagger U_{j+1}^{~} = \mathbb{1}$ that reduces the exponentially increasing spin degrees of freedom, i. e.,
\begin{equation}
    \begin{split}
        {\cal C}_{j+1}& =U_{j+1}^{~}\tilde{\cal C} _{j+1}U_{j+1}^\dagger\,,\\
        {\cal P}_{j+1}& =U_{j+1}^{~}\tilde{\cal P} _{j+1}U_{j+1}^\dagger\,.
    \end{split}
    \label{renp4short}
\end{equation}
This transformation process originates in the density matrix renormalization~\cite{DMRG1, DMRG2, DMRG3}. The computational cost of the optimized CTMRG code is $O(pq^3m^3)$ only. The details are provided in Appendix~\ref{ApA}.

The renormalization is accomplished by constructing the reduced density matrices at the iteration step $j$ from the $p$ extended $\tilde{\cal C}_{j+1}$
\begin{equation}
    \rho^\prime_{j+1} = {\rm Tr}^\prime (\tilde{\cal C}_{j+1}^p),
    \label{RDM}
\end{equation}
as the partial trace is taken over the environment degrees of freedom, i.e., all the spins except those on the cut depicted in Fig.~\ref{extension}. The isometry matrices $U_{j+1}$ are subsequently constructed from the reduced density matrix $\rho^\prime_{j+1}$. The isometries contain selected eigenvectors of $\rho^\prime_{j+1}$ corresponding to the $m$ largest eigenvalues. The integer $m\geq q$ controls the numerical accuracy and is referred to as the bond dimension or the number of states kept~\cite{DMRG1, DMRG2, DMRG3, DMRGsch}.

A remark on the Hermiticity of the reduced density matrix $\rho^\prime$ is needed at this stage.  For hyperbolic lattices $(p,4)$, $\rho^\prime$ is Hermitian for even integers $p=4,6,8,\dots$  From Fig.~\ref{extension}, we can see that the lattice can be divided into two identical halves only if $p$ is even, and each half of the lattice consists of $p/2$ corner transfer matrices. Their partial sum approximates the state $|\Psi\rangle = \sum'\tilde{\cal C}_{~}^{p/2}$ so that
\begin{equation}
 \rho^\prime = {\rm Tr}^\prime (\tilde{\cal C}_{~}^p)= {\rm Tr}^\prime (\tilde{\cal C}_{~}^{p/2}\tilde{\cal C}_{~}^{p/2}) = {\rm Tr}^\prime \vert \Psi \rangle\langle \Psi \vert.
\end{equation}
We keep the state normalized, i.e., $ \langle \Psi \vert \Psi \rangle=1$, for evaluating entanglement entropy correctly.

The odd values of $p$, lead to $\rho^\prime$ that is non-Hermitian, and we use such construction that involves the least asymmetry (and demand normalization $ \langle \Phi \vert \Psi \rangle=1$), i.e., 
\begin{equation}
 \rho^\prime = {\rm Tr}^\prime (\tilde{\cal C}_{~}^p)= {\rm Tr}^\prime (\tilde{\cal C}_{~}^{(p-1)/2}\tilde{\cal C}_{~}^{(p+1)/2}) = {\rm Tr}^\prime \vert \Psi \rangle\langle \Phi \vert.
\end{equation}

In Ref.~\onlinecite{triangular}, we studied hyperbolic lattices with non-Hermitian $\rho^\prime$ implementing three distinct approaches: (1) diagonalization of non-symmetric matrices, (2) singular value decomposition, and (3) diagonalization of symmetrized $\rho^\prime_{\rm sym}$. We encountered numerical instabilities in the cases (1) and (2),  which prevent the algorithm to converge. In case (3), the diagonalization of the symmetrized reduced density matrix $\rho^\prime_{\rm sym}= \tfrac{1}{2} (\vert \Psi \rangle\langle \Phi \vert + \vert \Phi \rangle\langle \Psi) \vert)$ always resulted in numerically stable solutions. This transform yields the optimal isometries $U$ with systematic and stable behavior of the algorithm. 

All numerical quantities, including the correct phase transition temperatures of the current vertex  TN completely coincide with the weight representation of TN~\cite{weightp4, weightpq} where the hyperbolic lattices $(p,4)$ always lead to the Hermitian (symmetric) construction of  $\rho^\prime$.  Alternatively, discovering a dual set of bi-orthonormal bases (corresponding to the left and right eigenvectors) can be a possible choice to reproduce the current results~\cite{BTMRG}.

\subsection{Thermodynamic quantities}
\label{ThQu}

After reaching the $k^{\rm th}$ iteration step, we calculate the partition function, magnetization, and other thermodynamic quantities. The partition function ${\cal Z}_{p,k}$ has the form
\begin{equation}
{\cal Z}_{p,k}
={\rm Tr}\, {\left({\cal C}_{k}^{p}\right)}=
\sum\limits_{\sigma_1\sigma_2\cdots\sigma_{{\cal N}_{p,k}}} \exp\left[-\frac{{\cal H}_p^{~}(\sigma_1\sigma_2\cdots\sigma_{{\cal N}_{p,k}})}{{k_{\rm B} T}}\right],
\label{pf}
\end{equation}
where ${\cal N}_{p,k}$ is the number of all spins on the $(p,4)$ lattice after $k$ iterations.

The spontaneous magnetization ${\cal M}_p = \langle {\cal S}_{{\sigma_c},\vartheta=0}\rangle_p$ (the order parameter in the bulk) is calculated as the mean value of the spin ${\sigma_c}$ placed in the lattice center (the bulk) to eliminate the boundary effects. In the thermodynamic limit ($k\to\infty$, i.e., ${\cal N}_{p,k}\to\infty$), we can evaluate the magnetization through the expression
\begin{equation}
{\cal M}_p = \lim\limits_{k\to\infty} \frac{\sum\limits_{\sigma_1\cdots\sigma_{{\cal N}_{p,k}}} {\cal S}_{{\sigma_c},0} \exp\left[-\frac{{\cal H}_p^{~}(\sigma_1\sigma_2\cdots\sigma_{{\cal N}_{p,k}})}{{k_{\rm B} T}}\right]}
{\sum\limits_{\sigma_1\sigma_2\cdots\sigma_{{\cal N}_{p,k}}} \exp\left[-\frac{{\cal H}_p^{~}(\sigma_1\sigma_2\cdots\sigma_{{\cal N}_{p,k}})}{{k_{\rm B} T}}\right]}.
\label{Mpq}
\end{equation}
This is equivalent to the expression
\begin{equation}
{\cal M}_p
=\frac{ {\rm Tr}\left[ {\cal I}_1\, {\cal P}_{\infty}^{4} {\cal C}_{\infty}^{4(p-3)} \right] }
      { {\rm Tr}\left[ {\cal V}\, {\cal P}_{\infty}^{4} {\cal C}_{\infty}^{4(p-3)} \right] }
\label{Mpq2}
\end{equation}
which is designed for CTMRG on the $(p,4)$ lattices. The central spin ${\cal S}_{{\sigma_c},\vartheta}$ is absorbed into a rank-4 impurity tensor
\begin{equation}
    {\cal I}_{abcd} = \sum\limits_{{\sigma}_c} {\cal S}_{{{\sigma}_c},\vartheta} Y_{{{\sigma}_c} a} Y_{{{\sigma}_c} b} Y_{{{\sigma}_c} c} Y_{{{\sigma}_c} d}\,.
    \label{Iabcd}
\end{equation}
Further details are discussed in Appendix~\ref{SpMg}.

It is also useful to examine the type of phase transition by analyzing the entanglement entropy. After constructing the reduced density matrix in the thermodynamic limit ($\rho_{k \to\infty}^\prime = \rho^\prime$), we compute the entanglement entropy
\begin{equation}
   {\cal E}_p = -{\rm Tr}\,(\rho^\prime\ln \rho^\prime)=-\sum\limits_{i=1}^m
    \omega_i\ln{\omega_i}
\label{Sv}
\end{equation}
where we consider the $m$ largest eigenvalues  $\omega_i$. For all hyperbolic lattices ($p>4$), we use a sufficiently large bond number $m$ to ensure that the truncation error is as small as $1 - \sum_{i=1}^m \omega_i < 10^{-30}$ (within the 128-bit numerical precision).

An additional remark on quantum entanglement entropy for classical spin systems is in place. We can ascribe classical spin models to quantum spin models using a quantum-classical correspondence~\cite{QCC, Chatelain}. The correspondence maps a $D$-dimensional quantum system onto a $(D+1)$-dimensional classical one due to the equivalence of their partition functions. The extra dimension in classical systems originates in the imaginary-time evolution via the Suzuki-Trotter expansion~\cite{Trotter, Suzuki1, Suzuki2}. 

In summary, the classical spin models in this work (studied on various $2D$ lattice surfaces) could be related to $1D$ quantum spin systems. For instance, the von Neumann entanglement entropy computed for the classical Ising model is identical to the associated quantum Ising model (after rescaling the transverse magnetic field $h \to h/h_{\rm c}$ and temperature $T \to T/T_{\rm c}$ in the corresponding classical model).

Finally, the free energy normalized per spin
\begin{equation}
    {\cal F}_{p,k}=-\frac{k_{\rm B} T}{ {\cal N}_{p,k}}\ln {\cal Z}_{p,k}
    \label{fe}
\end{equation}
can also be evaluated with high precision allowing us to take the second derivative with respect to temperature or the magnetic field~\cite{weightpq, triangular}. The free energy per spin is a well-defined (non-diverging quantity) in the thermodynamic limit. In Appendix~\ref{FreeEng}, we provide a derivation of the number of spins ${\cal N}_{p,k}$ and free energy $\ln {\cal F}_{p,k}$ we used in the numerical analysis.

\section{Results}
\label{results}

We selected the clock and $q$-state  Potts models because they exhibit first-order (discontinuous), second-order (continuous), and infinite-order Berezinskii-Kosterlitz-Thouless (BKT)~\cite{BKT1, BKT2, BKT3} transitions on the square lattice $(4,4)$. However, the hyperbolic geometry affects the above results since the Hausdorff dimension of the hyperbolic lattices is infinite. To verify the {\it vertex} representation of TN on the hyperbolic lattices, we classify the spin models by phase transition, as stated by the {\it weight} representation~\cite{Ueda1, weightpq, Baxter}.   We proceed with the following:
\begin{itemize}
    \item[{\bf A.}] Confirm the mean-field universality for the vertex TN on the pentagonal $(5,4)$ lattice;
    \item[{\bf B.}] Check the convergence of phase transition temperatures in the sequence of hyperbolic lattices $\{(p,4)\}_{p=5}^{\infty}$ to the Bethe lattice $(\infty,4)$;
    \item[{\bf C.}] Investigate the impact of higher-state spins ($q>2$) on the hyperbolic pentagonal lattice;
    \item[{\bf D.}] Analyze the multi-parametric properties of the free energy for the vertex TN on $(p,4)$ lattices.
\end{itemize}

\subsection{(5,4) lattice}

\begin{figure}[tb]
{\centering\includegraphics[width=\linewidth]{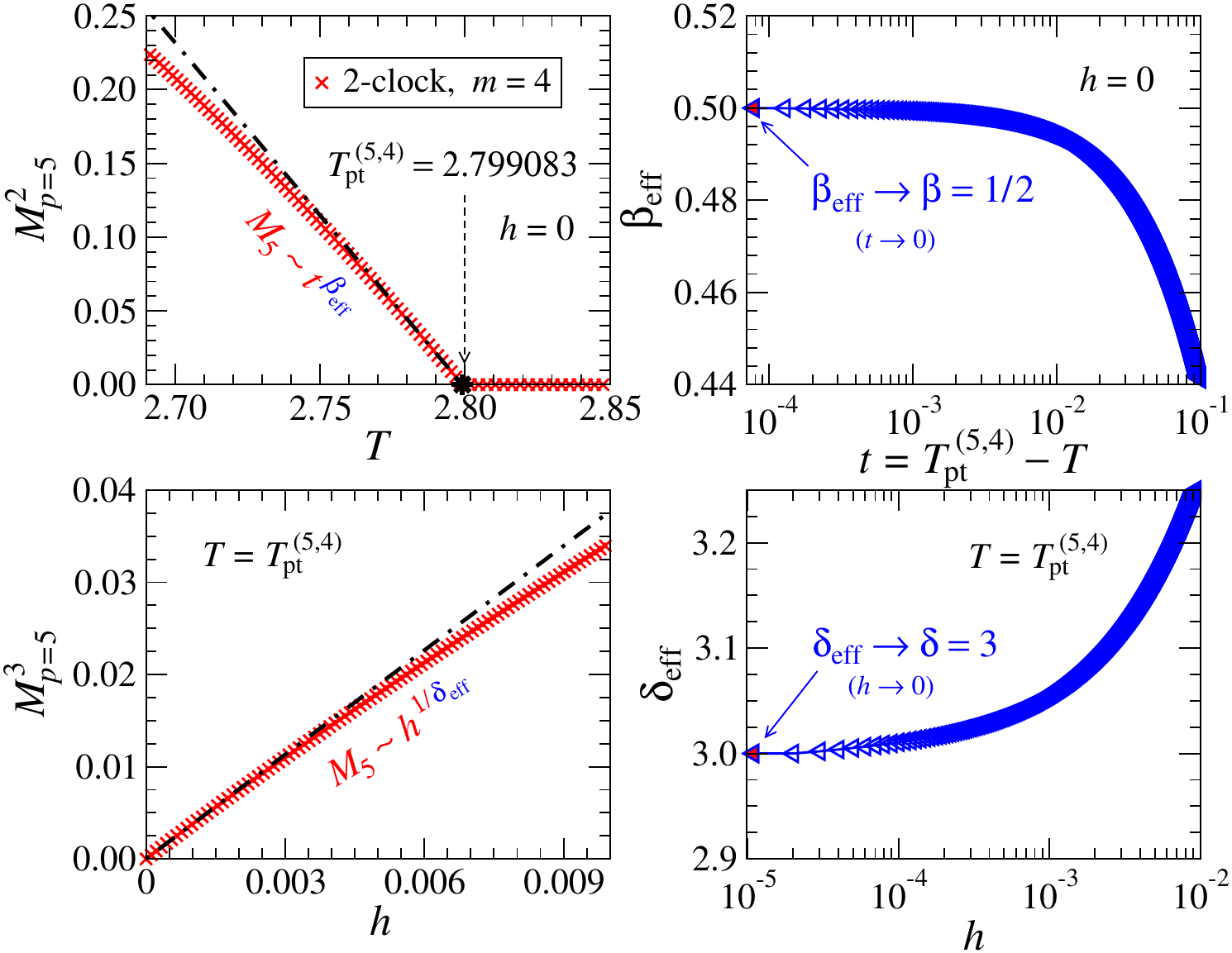}}
  \caption{The magnetization $M_5$ for the 2-state clock (Ising) model on the pentagonal lattice $(5,4)$ with the bond dimension as small as $m=4$. The dot-dashed lines (on the left) are guides for the eye exhibiting the linearity $M_5^2 \propto t$ in the limit $t\to0$ (at $h=0$) and $M_5^3 \propto h$ if $h\to0$ (at $t=0$). The mean-field exponents $\beta= \frac{1}{2}$ and $\delta=3$ are confirmed in the graphs on the right side.}
  \label{Mag_beta_delta}
\end{figure}

We begin with the Ising (2-state clock) model on the pentagonal $(5,4)$ lattice. In Fig.~\ref{Mag_beta_delta}, we analyze the Ising model and detect its phase-transition temperature $T_{\rm pt}^{(5,4)}$ in the limit $k\to\infty$. We evaluate the temperature dependence of spontaneous magnetization ${\cal M}_{p=5}$ at zero field magnetic field ($h=0$) and also the field-dependence of ${\cal M}_{5}$ at the phase-transition temperature $T_{\rm pt}^{(5,4)}$. We aim to accurately calculate phase-transition temperature $T_{\rm pt}^{(5,4)}$ and the associated magnetic exponents $\beta$ and $\delta$. We chose the phase-transition region where the vertex TN is subject to the lowest numerical accuracy due to the strongest correlations.

In the left top graph, we display the dependence of the squared spontaneous magnetization $M_5^2$ when linearly approaching the phase-transition temperature from the ordered ferromagnetic phase (this behavior is common for any $p \geq 5$ and will be shown later). Therefore, the spontaneous magnetization in the close vicinity of phase-transition temperature satisfies the scaling ${\cal M}_p \propto (T_{\rm pt}^{(p,4)} - T)^\beta$.

As an example, we evaluate the temperature-dependent effective exponent $\beta_{\rm eff}$ calculated just below the phase-transition temperature. Let $t=T_{\rm pt}^{(5,4)}-T$ be a small non-negative relative temperature such that $t \to 0$ at the phase transition. We calculate the effective exponent $\beta_{\rm eff}(t)$ at $h=0$ by taking the logarithmic derivative of the spontaneous magnetization ${\cal M}_{5} = const \cdot t^{\beta_{\rm eff}(t)}$. The accurate magnetization data is necessary to obtain a smooth dependence of $\beta_{\rm eff}$ on $t$ resulting in the phase transition temperature $T_{\rm pt}^{(5,4)}=2.799083$. We set a nonzero magnetic field $b$ imposed on the boundary to force the spontaneous symmetry breaking such that ${\cal M}_5>0$ in the ordered ferromagnetic phase $0\leq T < T_{\rm pt}^{(5,4)}$. If $\beta_{\rm eff}$ is  plotted as a function of $t$, we confirm the mean-field exponent~\cite{Baxter} $\beta\to\frac{1}{2}$ (for $p=5$) on the right top graph
\begin{equation}
    \beta = \lim\limits_{t\to0^+} \beta_{\rm eff}(t) = \lim\limits_{t\to0^+} \frac{\partial \ln {\cal M}_p(t,h=0)}{\partial \ln (t)} =  \frac{1}{2}\, .
\end{equation}

At the bottom of Fig.~\ref{Mag_beta_delta}, we show the mean-field exponent $\delta$ by taking the limit of the effective exponent $\delta_{\rm eff}(h)$ which features in the scaling relation ${\cal M}_5 \propto h^{1/\delta}$ measured at $T=T_{\rm pt}^{(5,4)}$ for $0\leq |h| \ll 1$ and the boundary field $b=0$. According to this relation, the cubed spontaneous magnetization ${\cal M}_p^3$ decreases to zero. In the bottom left graph, we display the linearizing dependence of the cubed spontaneous magnetization in direction $h\to0$ confirming the exact mean-field exponent
\begin{equation}
    \delta = \lim\limits_{h\to0} \delta_{\rm eff}(h) = \lim\limits_{h\to0}\left[\frac{\partial \ln {\cal M}_p(t=0,h)}{\partial \ln (h)}\right]^{-1} = 3\, ,
\end{equation}
as plotted in the right bottom graph on $(5,4)$. To summarize, as we approach the singularity, the linearity close to the phase transition of ${\cal M}_5^2$ vs $T$ and ${\cal M}_5^3$ vs $h$, respectively, yields the mean-field exponents $\beta=\frac{1}{2}$ and $\delta=3$. These results agree with the exact mean-field exponents and the weight representation of TN~\cite{MFU, triangular}.

In general, the mean-field universality, expressed via the exponents $\beta$ and $\delta$, holds for arbitrary hyperbolic lattices $(p,r)$ such that $(p-2)(r-2)>4$ since the Hausdorff dimension of such curved lattice surface is infinite~\cite{MFU}. In other words, the mean-field universality observed on the hyperbolic lattices~\cite{weightpq, triangular, weightp4} is not a consequence of the mean-field approximation because we have not introduced any such approximation into the CTMRG algorithm. Instead, the mean-field universality originates in the hyperbolic geometry itself since the Hausdorff dimension $d_H$ is infinite in the thermodynamic limit, i.e., if ${\cal N}_{p,k}\to\infty$. Recall that classical systems exceeding Hausdorff dimension $d_H=3$ display the mean-field universality~\cite{Baxter}.

Lastly, even at the phase transition, the correlation function decays exponentially~\cite{triangular} and the correlation length $\xi<1$ does not diverge~\cite{Iharagi, corr1, corr2}. Therefore, it is sufficient to perform the calculations with the bond dimension as small as $m=4$. An additional increase of $m$ does not affect the results since the improvement is negligible and of the order of numerical round-off errors~\cite{Vvsweight}. (This is not the case for the models on the Euclidean lattices. The numerical precision at the phase transition further improves and with increasing $m$ leads to the exactly known results of $T_{\rm c}^{(4,4)}=2/\ln(1+\sqrt{2})$, $\beta=\frac{1}{8}$, and $\delta=15$ only if the bond dimension grows accordingly, i.e., $m=q^k$.)

\subsection{({\itshape p},\,4) lattice}
\label{p4lat}

Now we check if the vertex representation of TN is correctly constructed. We calculate the phase-transition temperature of the Ising model for the Bethe lattice. The Bethe lattice with the coordination number $r=4$ can be accessed via the set of hyperbolic lattices and is equivalent to $(\infty,4)$ as also examined in Refs.~\onlinecite{weightpq, MFU, triangular}. The phase-transition temperature is exactly known~\cite{Baxter} to be  $T_{\rm pt}^{(\infty,4)}=2/\ln(2)$. The convergence to the Bethe lattice is common for both the quantum and the classical spin models we have studied in our earlier works by the weight representation only~\cite{AGquantum, Mi1, Mi2, weightp4}.

\begin{figure}[tb]
{\centering\includegraphics[width=\linewidth]{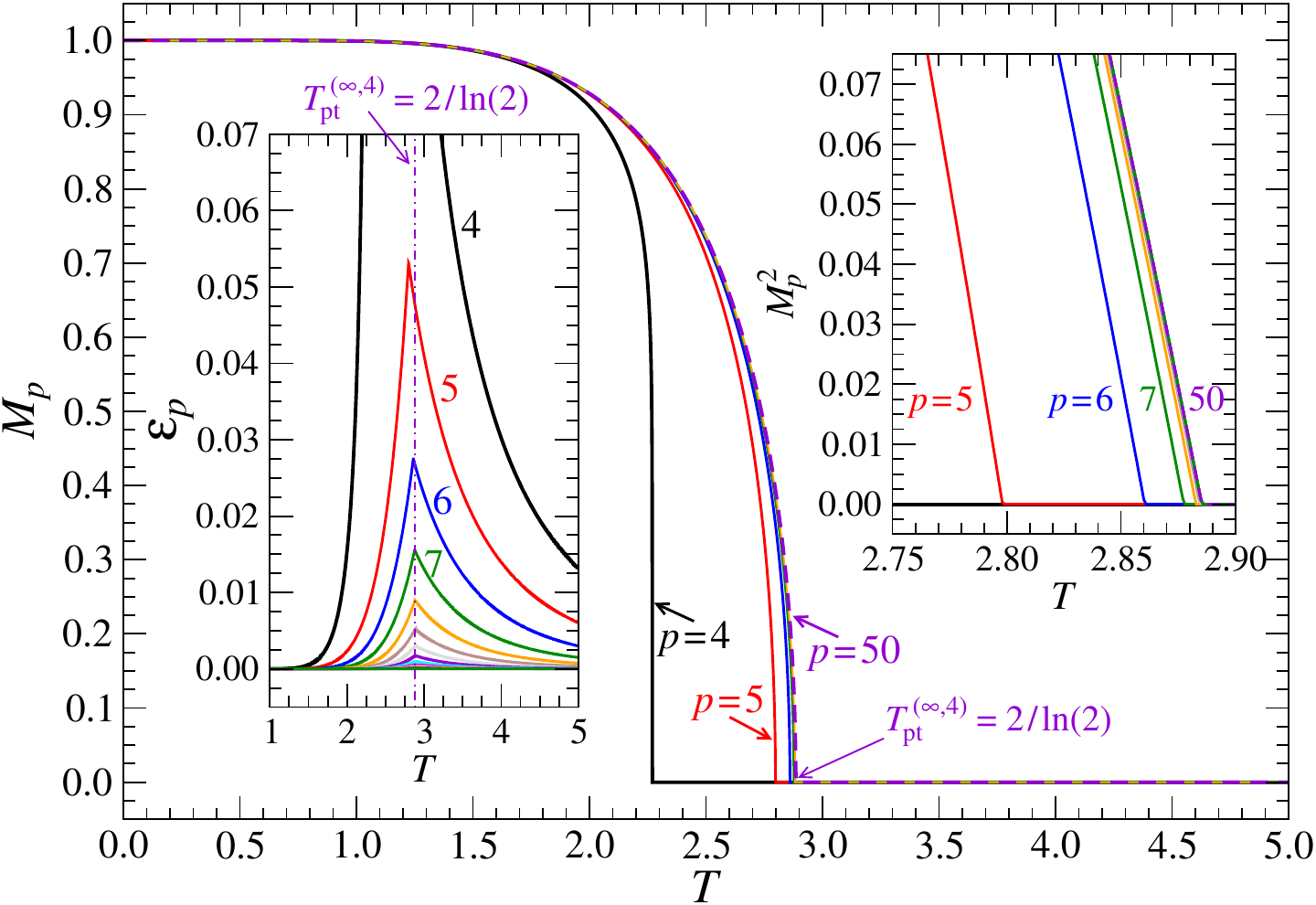}}
  \caption{The temperature dependence of the spontaneous magnetization towards the asymptotic Bethe lattice geometry $(\infty,4)$ with the bond dimension $m=4$. The left inset shows the fast asymptotic convergence of the entanglement entropy to the Bethe lattice. The right inset demonstrates the linearity, ${\cal M}^2_p \propto T$ leading to the mean-field exponent $\beta=\frac{1}{2}$.}
  \label{MpEp}
\end{figure}

In Fig.~\ref{MpEp} we plot multiple curves of spontaneous magnetization ${\cal M}_p$ by gradually expanding the size of the polygons $p=4,5,6,...,50$ corresponding to the lattices $(4,4)$, $(5,4)$, $(6,4),\cdots$, $(50,4)$. As $p$ increases, we observe a rapid asymptotic convergence to the phase-transition temperature of the Bethe lattice $(\infty,4)$. We notice that ${\cal M}_p$ for $p\gtrsim20$ are numerically indistinguishable from the Bethe lattice~\cite{weightp4}. The inset (on the right side) shows the linear decrease of ${\cal M}_p^2$ down to the phase-transition temperatures supporting the mean-field universality with $\beta=\frac{1}{2}$. 

The inset on the left in Fig.~\ref{MpEp} depicts the entanglement entropy ${\cal E}_p$ versus $T$ and $p$. We can clearly distinguish the $(p,4)$ lattices by the entanglement entropy because ${\cal E}_p$ decreases as $p$ grows, and its sharp (non-diverging) maximum refers to $T_{\rm pt}^{(p,4)}$. Surprisingly, the entanglement entropy remains sensitive to distinguish the lattice geometries with $p>20$. On the other hand, the magnetization and all other normalized thermodynamic quantities exhibit quantitatively indistinguishable bulk properties from $p\gtrsim20$.

\begin{figure}[tb]
{\centering\includegraphics[width=\linewidth]{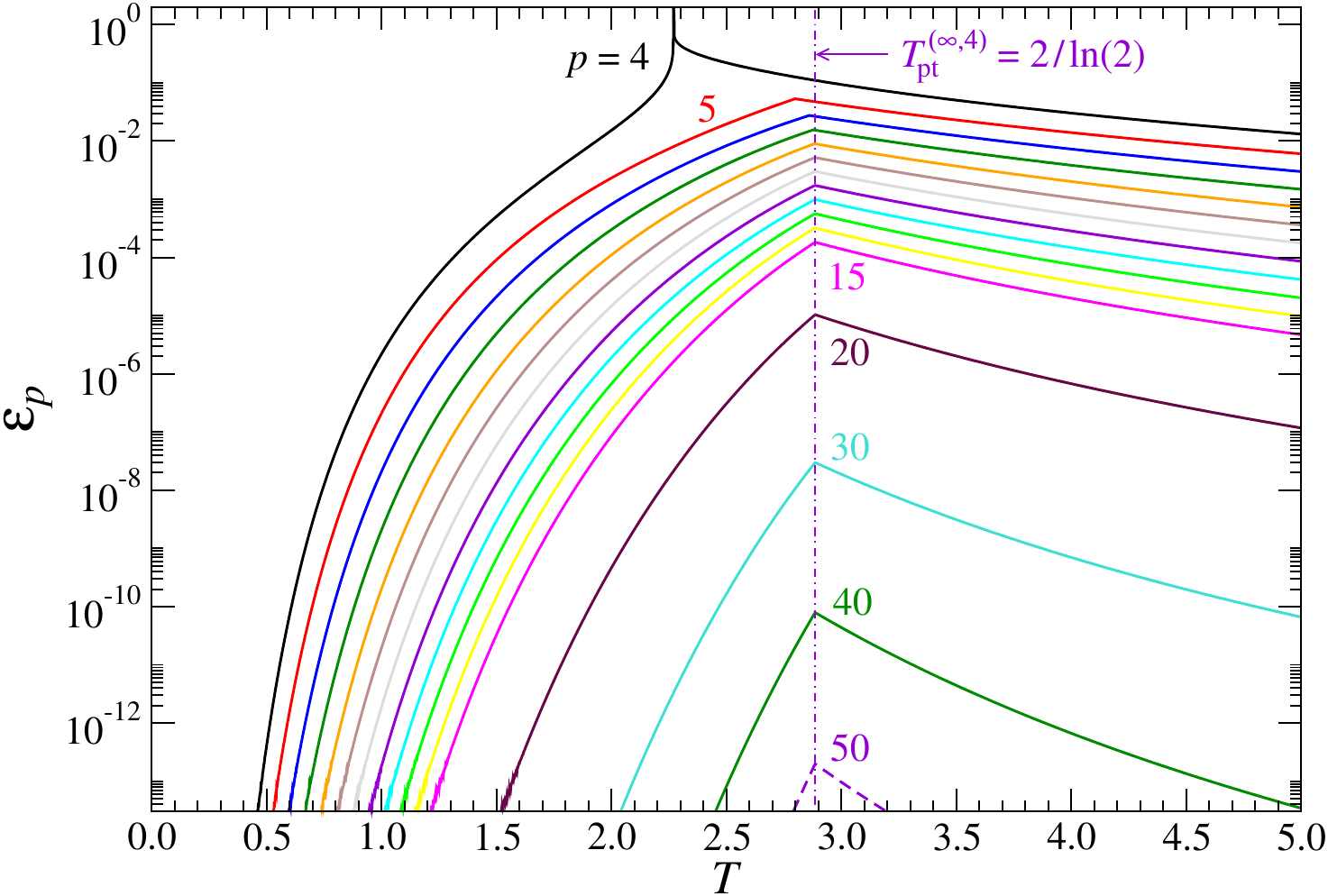}}
  \caption{The entanglement entropy versus temperature in lin-log scale for lattices $(p,4)$ where $4\leq p \leq 50$ with the bond dimension $m=4$. The asymptotic convergence of $T_{\rm pt}^{(p,4)}$ to the Bethe lattice is well visible on the sharp maxima of ${\cal E}_p$.}
  \label{Ep}
\end{figure}

Therefore, we show Fig.~\ref{Ep} to display the entanglement entropy in the semi-logarithmic scale to emphasize the differences among the various $(p,4)$ lattices. The decreasing peaks of ${\cal E}_p$ completely coincide with the transition temperatures in ${\cal M}_p$. In both cases, ${\cal M}_p$ and ${\cal E}_p$ approach the exact value $T_{\rm pt}^{(\infty,4)} = 2/\ln(2)$ equally rapidly on the Bethe lattice.

\begin{figure}[tb]
{\centering\includegraphics[width=\linewidth]{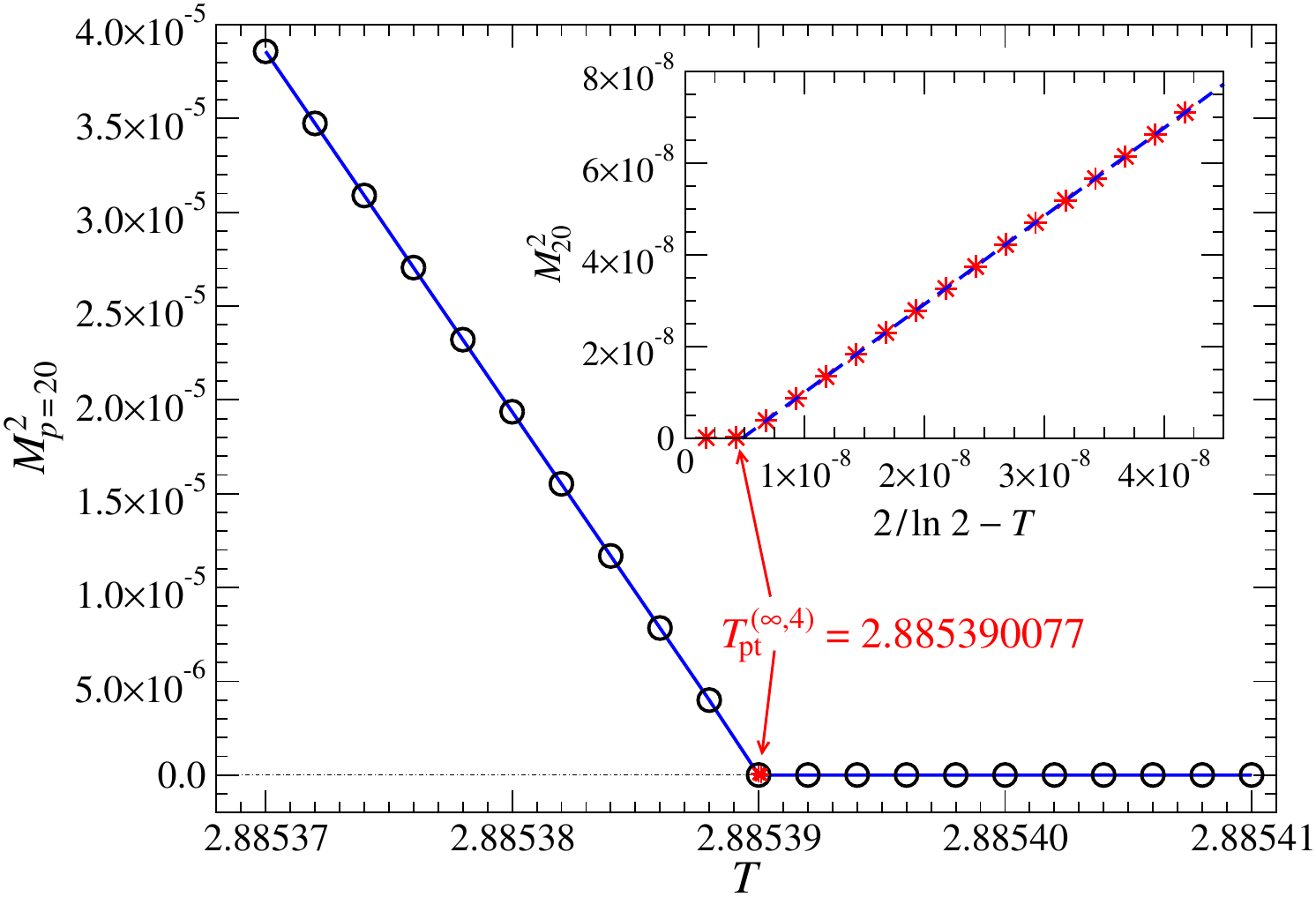}}
  \caption{Ising model ($q=2$ clock model) calculated on the hyperbolic $(20,4)$ lattice is numerically indistinguishable from the Bethe lattice. If fitting the mean-field exponent, we obtain $\beta =0.00002$ as can be read-off from the linear decrease of ${\cal M}_{20}^2$. The data points (asterisks) shown in the inset required $k > 10^{11}$ iteration cycles to converge at bond dimension $m=2$.}
  \label{M20}
\end{figure}

It is numerically more convenient to analyze $(20,4)$ than $(50,4)$ if treating the Bethe lattice $(\infty,4)$. Therefore, we plot Fig.~\ref{M20} with the squared magnetization ${\cal M}^2_{20}$ in the vicinity of phase-transition temperature $T_{\rm pt}^{(20,4)}$. The inset shows a highly detailed linear behavior of ${\cal M}^2_{20}$ versus shifted temperature $T_{\rm pt}^{(\infty,4)} - T$ and obtained the phase-transition temperature $T_{\rm pt}^{(20,4)} = T_{\rm pt}^{(\infty,4)} - 5\times 10^{-9} = 2.885390077$ for $m=2$ only. The exact Bethe lattice transition temperature~\cite{Baxter} occurs at $T_{\rm pt}^{(\infty,4)}=2/\ln(2)=2.885390082$. 

We set $m=2$ because of the fast decaying spectrum of the reduced density matrix $\rho'_j$. The eigenvalue spectrum of $\rho'_j$ decays faster with increasing $p$. Thus, such a low value of the bond dimension $m$ suffices and results in the small entanglement entropy referring to extremely weak correlations at the phase transition. Thus, the additional increase of the bond dimension to $m=4$ does not improve $T_{\rm pt}^{(20,4)}$. The inset highlights that the lattice ($20,4$) can retrieve the Bethe-lattice properties with a relative error for the temperature as small as $2\times10^{-7}\,$\%. Computing the data in the inset by the vertex-type TN required an extremely long time to reach the full convergence to the thermodynamic limit ($k>10^{11}$).

\subsection{Multi-state spin models}

Increasing the spin degrees of freedom enriches the variability of phase-transition types. For this purpose, we use the $q$-state clock and $q$-state Potts models. We select the range $2\leq q<10$ since these values are sufficient for covering the three basic types of phase transitions: the $1^{\rm st}$, $2^{\rm nd}$, and $\infty^{\rm th}$ orders. We test the multi-state spin models because the vertex TN displays the strongest long-range correlations leading to the lowest numerical efficiency.

\begin{figure}[tb]
{\centering\includegraphics[width=\linewidth]{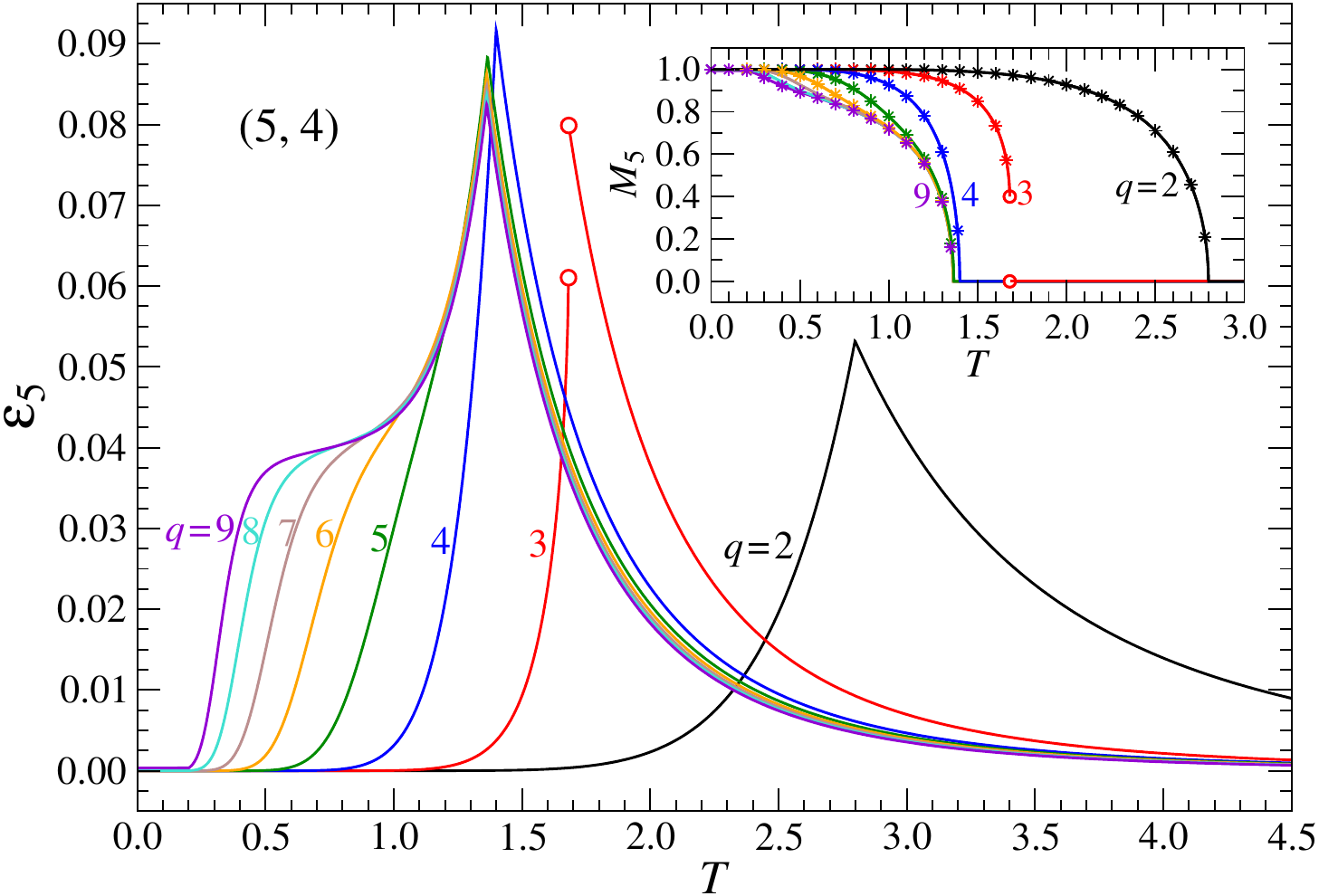}}
  \caption{The entanglement entropy ${\cal E}_{5}$ vs temperature for the $q$-state clock models on the lattice $(5,4)$ with the bond dimension $m=q^2$. The inset shows the temperature dependence of spontaneous magnetization $M_{5}$ for the $q$-state clock models (lines) that coincides with the results of the weight-type TN, shown by the symbols ($\ast$).}
  \label{E5M5_q-stateClock}
\end{figure}

In Fig.~\ref{E5M5_q-stateClock}, we classify the phase transitions by entanglement entropy ${\cal E}_{p=5}$ and spontaneous magnetization ${\cal M}_{p=5}$ in the $q$-state clock models on the pentagonal $(5,4)$ lattice (because there are no qualitative differences for $p>5$ and $q>9$ in such systems.) The infinite Hausdorff dimension of the hyperbolic structure of TN still affects the types of phase transitions and differs from those on the Euclidean lattice $(4,4)$.

We detect the second-order phase transitions (belonging to the mean-field universality class) for $q=2$ (the Ising model) and $q=4$ (two decoupled Ising models). At $q=3$, however, we observe a phase transition of the first order although the $3$-state clock model exhibits the second-order transition on the square lattice belonging to the $3$-state Potts universality class~\cite{FYWu}. The first-order discontinuity occurs on $(5,4)$ because the $3$-state Potts universality in dimensions $d\geq3$ has to be of the first order.

We point out the non-diverging entropy ${\cal E}_5$ in the clock model for $q\geq5$, as plotted in Fig.~\ref{E5M5_q-stateClock}. Recall that ${\cal E}_4$ for the $q$-state clock models with $q\geq2$ on the Euclidean ($4,4$) lattices logarithmically diverges at phase transitions~\cite{MP, Ueda, CentrCharge}. On the other hand, phase transitions on the hyperbolic lattices are always non-critical~\cite{MFU, triangular} since the correlation length $\xi$ does not diverge~\cite{Iharagi}, reaches a maximal sharp peak and is small ($\xi<1$).

Entanglement entropy at phase transition for $h=b=0$ diverges logarithmically (${\cal E}_4 \propto \frac{1}{12}\ln k$) on the Euclidean two-dimensional lattices~\cite{CentrCharge}, while ${\cal E}_{p\geq5}$ is always finite on the hyperbolic lattices $0 < {\cal E}_{{p\geq5}} \ll 1$. The correlations decay exponentially if measured between the bulk and the lattice boundary. However, the correlations on the lattice boundary always decay as a power law, and the system is critical on the hyperbolic boundary~\cite{TNAds1}. Comparing the qualitative similarity between the non-diverging ${\cal E}_5$ with the logarithmically diverging ${\cal E}_4$ on the square lattice~\cite{MP}, we can conjecture a non-critical BKT-like phase that exists in bulk only. However, no BKT phase transition is present for $p\geq5$.

Spontaneous magnetization ${\cal M}_5$ for the clock model is plotted in the inset of Fig.~\ref{E5M5_q-stateClock}. It is nonzero in the ordered ferromagnetic ($2\leq q\leq4$)  phase, zero in the disordered phase, and drops to zero at the phase transition. The phase-transition discontinuity at $q=3$ confirms the $1^{\rm st}$ transition as the same discontinuity in ${\cal E}_5$. For $q\geq5$ at $T<T_{\rm pt}^{(5,4)}$, the spontaneous magnetization $M_5$ exhibits a small ripple indicating the presence of an intermediate BKT phase which separates the low-temperature ferromagnetic phase from the high-temperature paramagnetic phase. Tiny magnetization ripples in the intermediate BKT region are typical in ${\cal M}_4$ on the square $(4,4)$ lattice.

In the inset of Fig.~\ref{E5M5_q-stateClock}, we compare ${\cal M}_5$ in the current vertex-type TN study (full lines) with our earlier work on the weight-type TN (asterisks). The data fully coincides, including the free energy and other derived thermodynamic properties, such as internal energy, specific heat, and susceptibility (not shown). There is, however, a significant difference in the entanglement entropy between the vertex and weight representations of TN which deserves deeper analysis and is to be thoroughly studied elsewhere~\cite{Vvsweight}.

\begin{figure}[tb]
{\centering\includegraphics[width=\linewidth]{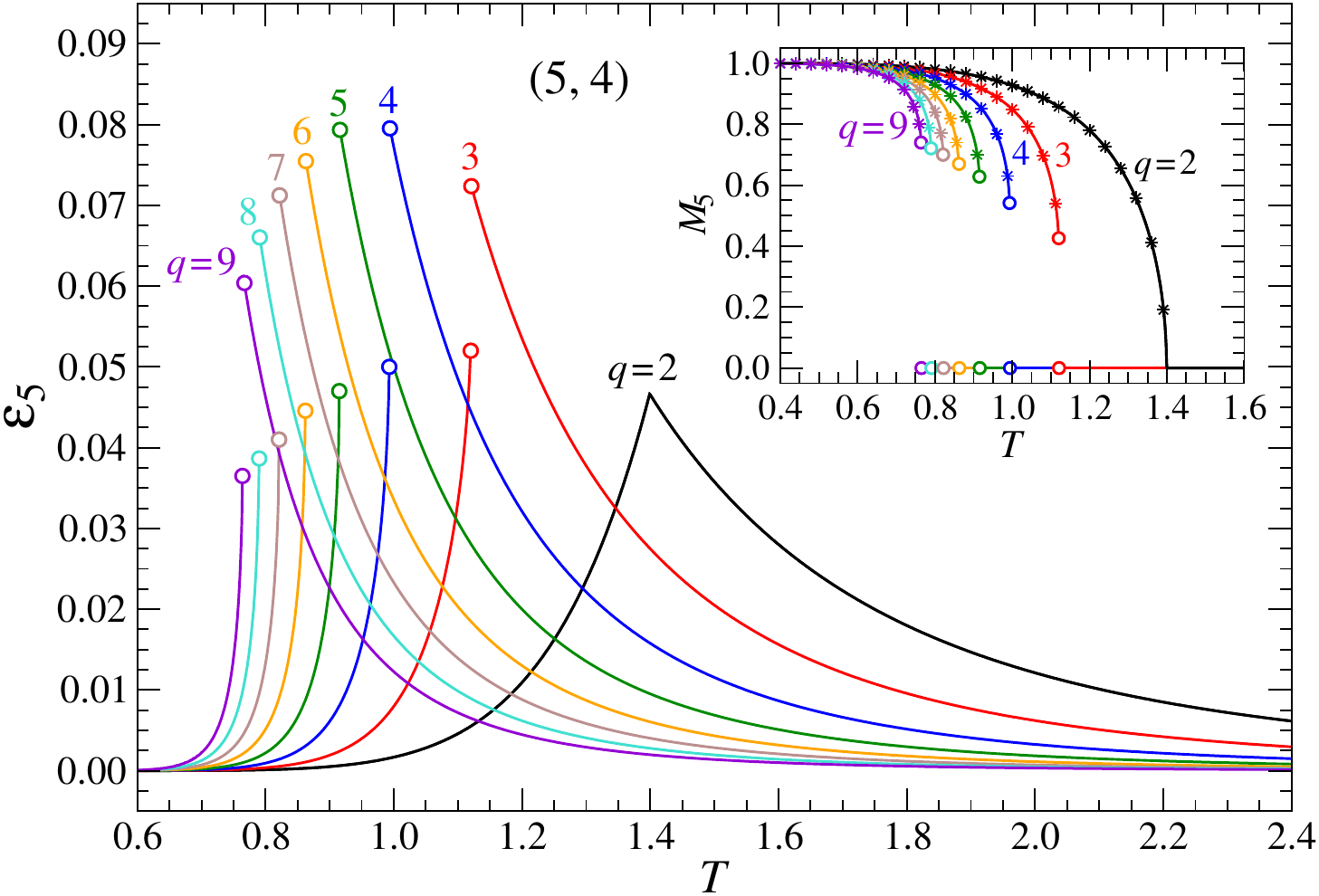}}
  \caption{The entanglement entropy ${\cal E}_5$ vs temperature for the $q$-state Potts models on $(5,4)$ lattice with the bond dimension $m=q^2$. We emphasize the emergence of the first-order transition for $q\geq 3$ on the hyperbolic lattices. The circles denote the phase-transition temperatures where the curves become discontinuous. Inset shows the equivalence of spontaneous magnetization ${\cal M}_5$ between the vertex TNs (lines) and the weight TNs ($\ast$).}
  \label{E5M5_q-statePotts}
\end{figure}

Figure~\ref{E5M5_q-statePotts} displays the entanglement entropy for the $q$-state Potts model on the ($5,4$) lattice. The $2$-Potts and Ising models are identical after rescaling $J\to2J$. The $q$-state Potts models on the square $(4,4)$ lattice~\cite{MP} exhibit the second-order phase transition $q=2,3,4$ whereas the first-order (discontinuous) transition is present for $q\geq5$. On the lattices with dimension $d\geq3$ (including the hyperbolic lattices with $d\to\infty$), the first-order transition occurs for $q \geq 3$~\cite{FYWu, Baxter}. The discontinuous jumps refers to the $1^{\rm st}$ order transition in ${\cal E}_5$ and ${\cal M}_5$ (inset) at phase-transition temperatures $T_{\rm pt}^{(5,4)}$.

\subsection{Free Energy}

When analyzing the free energy on $(p,4)$, as mentioned in Sec.~\ref{ThQu}, every vertex of TN contributes to ${\cal F}$, mainly those on the lattice boundary. Moreover, ${\cal F}$ depends on the spin model (specified by spin type ${\cal S}$), temperature $T$, interaction $J$, magnetic field $h$ and the boundary field $b$, lattice geometry $(p,r)$ and the number of spin states $q$. To check whether the vertex-type TN has been correctly constructed, we examine zero-temperature and high-temperature limits for which ${\cal F}$ acquires analytical expressions, see Appendix~\ref{FEas}.

\begin{figure}[tb]
{\centering\includegraphics[width=\linewidth]{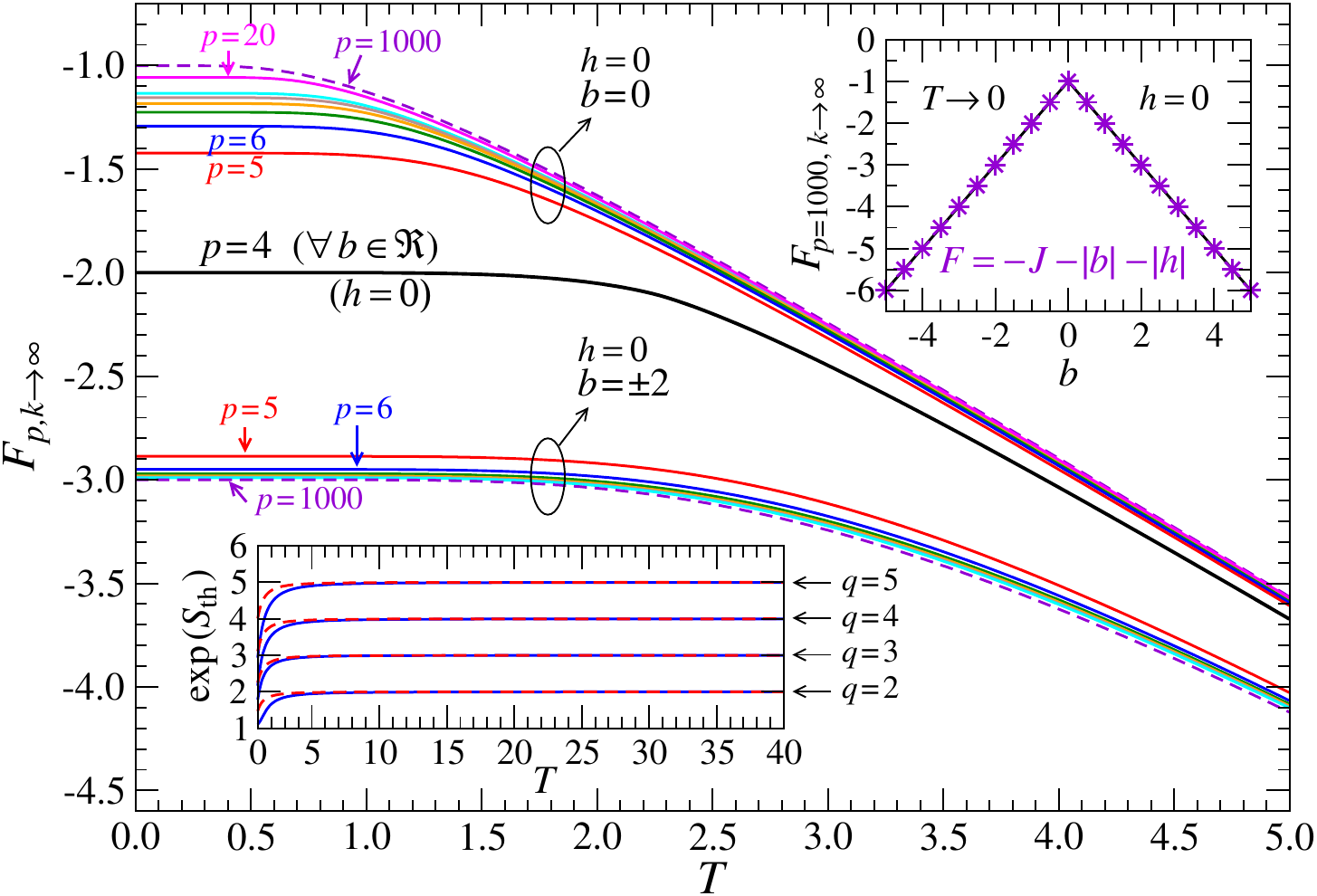}}
  \caption{The temperature dependence of the free energy for the Ising model ($q=2$) on $(p,4)$ lattices when imposing the boundary field $b=0$ and $b=\pm2$ (at $h=0$). Calculated for the bond dimension $m=q^2$. The free energy does not depend on the boundary field $b$ on the square ($4,4$) lattice. The upper inset depicts a strictly linear dependence of the free energy on $b$ and $h$ at $T\to0$ if studied on the Bethe lattice. The lower inset confirms the high-temperature regime of the thermodynamic entropy $S_{\rm th} \to \ln(q)$. The full and dashed lines correspond to the $q$-state clock and Potts models, respectively.}
  \label{FreeEnergy}
\end{figure}

Recalling the full Hamiltonian in Eq.~\eqref{H}, we primarily examine the response of the boundary magnetic field $b$ on the entire system described by ${\cal F}$. Because the number of boundary spins on the hyperbolic lattices is larger than the number of the remaining inner spins for any $k$ (even in the thermodynamic limit). The field $b$ significantly affects the bulk properties and suppresses the existence of phase transitions in the bulk. This is certainly true if extracting the specific heat after taking the second derivative of the free energy with respect to temperature~\cite{weightpq}. However, this is not the case for the Euclidean lattices where the ratio between the boundary and inner spins tends to zero in the thermodynamic limit.

In Fig.~\ref{FreeEnergy} we plot the functional dependence of the free energy on the temperature on $(p,4)$ lattices with $4 \leq p \leq 1000$ parameterized by the boundary field $b$. Setting $b=0$ corresponds to the free boundary conditions while the fixed boundary conditions apply for nonzero $b$. The larger the field $b$ (in the absolute value), the stronger the impact on the lattice system is observed. We have analytically derived the low- and high-temperature limits of free energy in the vertex TN that satisfy the numerical results shown in the graph (see App.~\ref{FEas}).

In the low-temperature limit, free energy per spin acquires only negative values. For instance, ${\cal F}_{p=4,k\to\infty}=-2$ on the square $(4,4)$ lattice at $b=0$, whereas on the Bethe lattice we get ${\cal F}_{p\to\infty,k\to\infty}=-1$ for $b=0$ and ${\cal F}_{p\to\infty,k\to\infty}=-3$ at $b=\pm2$, for all cases we used $h=0$ and $T < 0.01$. We simulated the Bethe lattice $(1000,4)$ at low temperatures after setting the 128-bit numerical precision (the quartic precision up to 34 digits).

In the top right inset of Fig.~\ref{FreeEnergy}, we show a linear dependence of the free energy versus the boundary magnetic field $b$. To prove the correctness, we have derived the free energy at zero temperature on the square lattice ($4,4$)
\begin{equation}
    \lim\limits_{T\to0} {\cal F}({\cal S},T,J,h,b,p=4,r,q)= -2J -|h|.
    \label{LowTempF}
\end{equation}
On the Bethe lattice ($\infty,4$), we get
\begin{equation}
    \lim\limits_{\substack{T\to0\\p\to\infty}} {\cal F}({\cal S},T,J,h,b,p,r,q)= -J -|h| -|b|.
    \label{LowTempF}
\end{equation}
Concise analytic derivation of ${\cal F}$ is given in Appendix~\ref{FEzero}, agreeing with the data shown in the graph.

The free-energy slopes for $q=2$ become identical in the high-temperature limit, as depicted in Fig.~\ref{FreeEnergy}. This feature is robust for a wide range of parameters. Taking the limit $T\to\infty$, the thermodynamic entropy $S_{\rm th}=-{\rm d}{\cal F}/{\rm d}T$ for the vertex representation of TN gives (for details, see Appendix~\ref{FEinf})
\begin{equation}
    \lim\limits_{T\to\infty} -\frac{\rm d} {{\rm d}T}\, 
{\cal F}({\cal S},T,J,h,b,p,r,q)=\ln\,(q)\, ,    
\label{Sth}
\end{equation}
which agrees with the results of the weight-type algorithm~\cite{weightpq, Axelrod}. We have numerically verified that the asymptotics of ${\cal F}({\cal S}, T, J,h,b,p,r,q)$ does not depend on the spin model ${\cal S}$, the spin-spin interaction $J$, both magnetic fields $h$ and $b$, or the lattice geometry ($p,r$) satisfying $(p-2)(r-2)\geq4$. To emphasize, we verified the $(p,4)$ lattices by the vertex-type TN, and the remaining lattices $(p,r)$ with $r\neq4$ have been confirmed by the weight TN calculations, see Ref.~\onlinecite{weightpq}.

We have numerically verified (not shown) that the free energies for the vertex and weight representations are identical for arbitrary $q$-state clock and Potts models on any $(p,4)$ lattices, provided that $h=b=0$. If imposing nonzero magnetic fields $h$ and $b$, the free energies calculated for the vertex and weight TNs differ by construction because the hyperbolic structure of the lattices leads to unequal numbers of boundary spins for any $k$. 

Contrary to the lattices in the Euclidean spaces, the magnetic field $b$ applied to the boundary spins on the hyperbolic lattices strongly affects global physical properties even in the thermodynamic limit. It is so because the number of boundary spins prevails over the remaining spins~\cite{MFU}. Hence, no phase transition exists on the entire lattice (no spontaneously broken symmetry leads to the ordered ferromagnetic phase). In other words, the specific heat $c = -T \partial^2 {\cal F}/\partial T^2$ on hyperbolic lattices does not result in a diverging peak at phase transition $T_{\rm pt}^{(p,4)}$. Instead, a broadened maximum at much lower temperatures occurs~\cite{weightpq}. The phase transition on the hyperbolic lattices exists in the deep bulk only, i.e., infinitely far from the boundaries, as stated for the Bethe lattice in Ref.~\onlinecite{Baxter}.

\section{Conclusion}
\label{Concl}
We proposed the vertex representation of tensor networks modeled on the specified class of hyperbolic lattices $(p,r)$ with the fixed coordination number $r=4$. The extension for arbitrary $r$ is straightforward. We have tested the vertex TN on the exactly solvable Bethe lattice to confirm the correctness of the $(p,4)$ constructions, and, at the same time, we confirmed that the vertex TN exhibits identical results to the weight-type TN~\cite{weightpq}. We did not perform calculations with fixed $p$ while varying $r$ since we conjecture the equivalent behavior with the weight representation, i.e., as $r$ increases, the phase transition temperature linearly increases ($T_{\rm pt}^{(p,r)}\propto r$ for $r>p$ as in our earlier work~\cite{weightpq}).

We verified the mean-field nature of the vertex-type TN as expected on the hyperbolic geometry by calculating the critical exponents $\beta = \frac{1}{2}$ and $\delta = 3$. In analogy to the weight-type TN~\cite{weightpq}, we have again observed a fast convergence of the sequence of hyperbolic lattices $\{(p,4)\}_{p=5}^{\infty}$ to the Bethe lattice $(\infty,4)$. We found out that the hyperbolic lattice $(20,4)$ (i.e., $p\geq20$) becomes numerically indistinguishable from the exact solution on the Bethe lattice~\cite{Baxter}. We compared the phase-transition temperature of the Ising model on the Bethe lattice showing the agreement with the nine valid digits. The uniqueness of entanglement entropy lies in the sensitivity to detect differences in the lattice geometry $(p,r)$, while the other quantities saturate and become insensitive. 

We have tested the vertex TN on the $q$-state spin models. On the hyperbolic lattices, the higher-spin models showed qualitative agreement with the models in the three or higher dimensions when dealing with the first- and second-order phase transitions. The BKT-like phase seems to be present deeply in the bulk only. All thermodynamic quantities on the hyperbolic lattices are non-critical, i.e., the correlation length, specific heat, and entanglement entropy do not diverge at phase transitions, including the correlation function which decays exponentially at phase transition along geodesics~\cite{MFU, weightp4, Iharagi, corr1, corr2}. We have observed an exceptional behavior in the entanglement entropy that deserves special attention and will be published soon~\cite{Vvsweight}.

If comparing the current results via the {\it vertex} representation of the TN with the earlier {\it weight} representation of TN~\cite{weightp4, weightpq}, we confirm the full agreement between the two representations. Moreover, our results for the Ising model on the (5,4) lattice also agree with Monte Carlo (MC) simulations on a $(5,5)$ hyperbolic lattice~\cite{QMC}, where the authors also confirm the mean-field universality specifying phase-transition temperature.

The numerical accuracy using the CTMRG method prevails over the MC simulations for 2D lattices. While the outermost spin layers must be removed using MC, the CTMRG algorithm neglects boundary effects describing the bulk properties. Moreover, the authors~\cite{QMC} numerically calculated the mean-field universality exponent $\beta = 0.46 \pm  0.5$, whereas we achieved the exponent $\beta = 0.500002$ almost reaching the exact value $\beta=\tfrac{1}{2}$. The deviations in MC arose due to finite-size scaling and the necessity to subtract a couple of boundary layers which significantly affect the bulk properties. In contrast, the most relevant errors in CTMRG originate from the round-off errors due to the large number of iterations (since the hyperbolic systems are off-critical and setting the bond dimensions $m\approx q^2$ is sufficient due to the exponential decay of density matrix eigenvalues~\cite{Vvsweight}).  

Having compared the most similar hyperbolic lattices (5, 5) calculated by MC and (5, 4) by CTMRG, we see that MC yields the phase-ransition temperature $T_c/J = 3.93 \pm 0.03$ on the (5, 5) lattice while CTMRG results in the phase transition $T_c/J = 2.799083$ for (5, 4) lattice. The phase-transition temperature by CTMRG can be further improved if more iterations are taken, as demonstrated on the Bethe lattice in Fig.~\ref{M20}. Away from the phase transition, the numerical accuracy of all physical quantities reaches the machine precision.  On the other hand, the MC simulations have been successfully applied to in the 3D hyperbolic space,  and we hope to make improvements in the future). We notice that CTMRG is not accurate for the 3D cubic lattice where substantially large bond dimensions are necessary~\cite{3DCTMRG}.

We aimed to develop the vertex representation because its implementation can be used for quantum systems by PEPS~\cite{PEPS} construction on the hyperbolic lattices. The implementation is straightforward by adding an extra physical index $x$ into the optimized tensors which represents the PEPS on the hyperbolic lattices such that
\begin{equation*}
    \begin{split}
        |\psi\rangle_{(p,4)} & = \hspace{-0.1cm} \sum\limits_{x_1^{~} x_2^{~}\dots x_{\cal N}^{~}=0}^{q^{\cal N}} \hspace{-0.2cm} {\rm Tr}\,\left( [{\cal V}]^{x_1^{~}}[{\cal V}]^{x_2^{~}}\cdots[{\cal V}]^{x_{\cal N}^{~}}\right)|x_1^{~} x_2^{~}\dots x_{\cal N}^{~}\rangle\\
        &= \hspace{-0.1cm} \sum\limits_{x_{\cal V}^{~}=0}^{q^{p}} \sum\limits_{x_{\cal P}^{~}=0}^{q^{2p}} \hspace{-0.1cm} \sum\limits_{x_{\cal C}^{~}=0}^{q^{p(2p-7)}} \hspace{-0.2cm} {\rm Tr}\left( \left[ {\cal V}\right]^{x_{\cal V}^{~}}_{~}\left[{\cal P} \right]^{x_{\cal P}^{~}}_{~}\left[{\cal C} \right]^{x_{\cal C}^{~}}_{~} \right)|x_{\cal V}^{~} x_{\cal P}^{~} x_{\cal C}^{~} \rangle.
    \end{split}
\end{equation*}
The trace goes over the multiple tensors ${\cal V}$, ${\cal P}$, and ${\cal C}$ that are combined to form ($p,4$) lattices, as depicted in Figs.~\ref{C1-C2}, \ref{extension}, and \ref{Mag_CTMRG}. In the CTMRG language, the PEPS requires the following extension of the tensor indices
\begin{equation*}
    \begin{split}
        [{\cal V}]_{abcd} &\to [{\cal V}]^{x_{\cal V}^{~}}_{abcd} \\
        [{\cal P}]_{abc\phantom{d}} &\to [{\cal P}]^{x_{\cal P}^{~}}_{abc} \\
        [{\cal C}]_{ab\phantom{cd}} &\to [{\cal C}]^{x_{\cal C}^{~}}_{ab} 
    \end{split}
\end{equation*}
The number of the physical spin indices $x_{\cal V}^{~}$, $x_{\cal P}^{~}$, $x_{\cal C}^{~}$ depends on the TN structure that is $p$, $2p$, $p(2p-7)$, respectively, in accord with the right-hand side of Fig.~\ref{Mag_CTMRG}. Such a construction is inevitable for the calculation of $\langle\psi|{\cal H}|\psi\rangle_{(p,4)}$ and other observables on hyperbolic lattices. 


\section{Acknowledgments}
This work was partially funded by Agent\'{u}ra
pre Podporu V\'{y}skumu a V\'{y}voja (No. APVV-20-0150), Vedeck\'{a} Grantov\'{a} Agent\'{u}ra M\v{S}VVa\v{S} SR and SAV (VEGA Grant No. 2/0156/22).

\appendix 

\section{Extension and renormalization}
\label{ApA}

\begin{figure}[tb]
{\centering\includegraphics[width=0.462\linewidth]{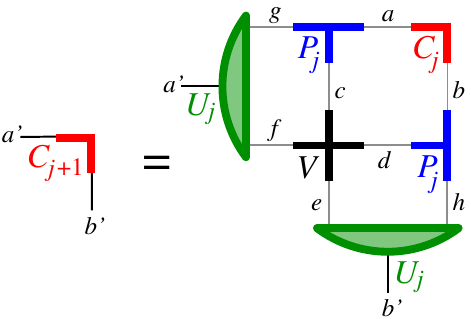}}\qquad 
{\centering\includegraphics[width=0.462\linewidth]{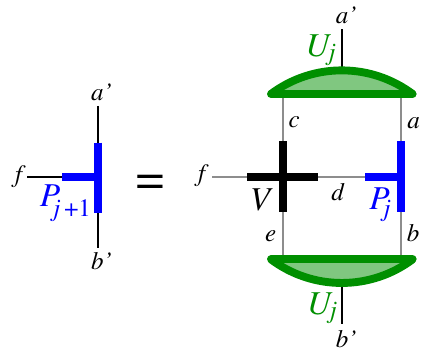}}\\
\medskip\medskip
{\centering\includegraphics[width=0.462\linewidth]{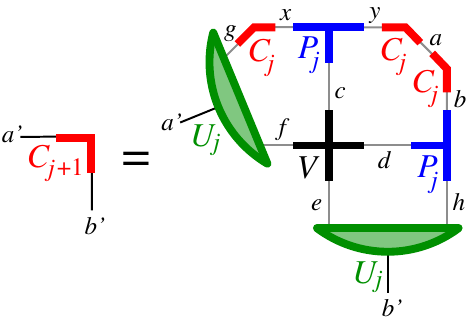}}\qquad 
{\centering\includegraphics[width=0.462\linewidth]{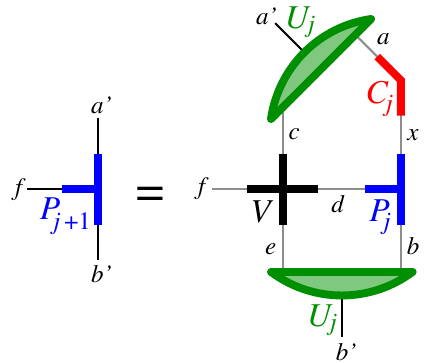}}
  \caption{Graphical representation of the extension and renormalization processes joined together, as separately defined in Eqs.~\eqref{extp4short} and \eqref{renp4short}, respectively. The upper panel depicts Eqs.~\eqref{ext44} and \eqref{renp4} for the square $(4,4)$ lattice and the lower panel illustrates Eqs.~\eqref{ext54} and \eqref{renp4} for the hyperbolic pentagonal $(5,4)$ lattice. The isometry ${\cal U}$ is associated with the reduced density matrix in Eq.~\eqref{RDM}.}
  \label{renormalization}
\end{figure}

Here we describe the extension process using both the bond indices $a,b,c$ and the iteration step index $j$ in the tensors ${[{\cal P}_j]}_{abc}$ and ${[{\cal C}_j]}_{ab}$. Reviewing the extension part for the Euclidean square $(4,4)$ lattice means considering the following construction (see the upper panel in Fig.~\ref{renormalization})
\begin{equation}
\begin{split}
&{[\tilde{\cal C}_{j+1}]}_{ehfg}=\sum\limits_{\substack{abcd\\}}{[{\cal V}]}_{cdef}\,{[{\cal P}_j]}_{gca} {[{\cal P}_j]}_{bdh}\,{[{\cal C}_{j}]}_{ab}^{~} \,,\\
&{[\tilde{\cal P}_{j+1}]}_{ebcaf}=\sum\limits_{d}{[{\cal V}]}_{cdef}\,{[{\cal P}_{j}]}_{adb}^{~}\,.
\end{split} 
\label{ext44}
\end{equation}
On the hyperbolic pentagonal $(5,4)$ lattice, we get (see the lower panel in Fig.~\ref{renormalization})
\begin{equation}
\begin{split}
&{[\tilde{\cal C}_{j+1}]}_{ehfg}=\sum\limits_{\substack{abcd\\}}{[{\cal V}]}_{cdef}{[{\cal P}_j]}_{xcy} {[{\cal P}_j]}_{bdh}{[{\cal C}_{j}]}_{ab}^{~} {[{\cal C}_{j}]}_{ya}^{~} {[{\cal C}_{j}]}_{gx}^{~} ,\\
&{[\tilde{\cal P}_{j+1}]}_{ebcaf}=\sum\limits_{xd}{[{\cal V}]}_{cdef}{[{\cal P}_{j}]}_{xdb}^{~}{[{\cal C}_{j}]}_{ax}^{~}.
\end{split} 
\label{ext54}
\end{equation}

Thus extended rank-$4$ tensor ${[\tilde{\cal C}_{j+1}]}_{ehfg}$ and the rank-$5$ tensor ${[\tilde{\cal P}_{j+1}]}_{ebcaf}$ enter the renormalization transformation. After applying the isometry $U_{j+1}$, we obtain
\begin{equation}
\begin{split}
&{[{\cal C}_{j+1}]}_{a^\prime b^\prime}=\sum\limits_{efgh} {[{\cal U}_{j+1} ]}_{a^\prime gf} {[\tilde{\cal C}_{j+1}]}_{ehfg}\, {[{\cal U}_{j+1}^\dagger]}_{he\,b^\prime}\,, \\
&{[{\cal P}_{j+1}]}_{a^\prime fb^\prime}=\sum\limits_{abce}{[{\cal U}_{j+1}]}_{a^\prime ac}{[\tilde{\cal P}_{j+1}]}_{ebcaf}\,{[{\cal U}_{j+1}^\dagger]}_{be\,b^\prime}\,,
\end{split}
\label{renp4}
\end{equation}
and the extended and renormalized tensors transform back to the rank-$2$ tensor ${[{\cal C}_{j+1}]}_{a'b'}$ and the rank-$3$ tensor ${[{\cal P}_{j+1}]}_{a'fb'}$, regardless of the lattice geometry $(p,4)$, as graphically depicted in Fig.~\ref{renormalization}.

The reduced density matrix $\rho_{j+1}^{\prime}$  defined in Eq.~\ref{RDM} is formed for the extended $\tilde{\cal C}_{j+1}$ where $j$ enumerates the iterations $j=1,2,3,\dots,k$. For brevity, let $j'=j+1$. The dimension of $\rho_{j'}^{\prime}$ would grow exponentially as $q^{j+1}$ if the tensors were not renormalized. Recall that $\rho_{j'}^{\prime}$ is defined along the geodesic between any two adjacent $\tilde{\cal C}_{j'}$, see Fig.~\ref{extension}. For instance, the reduced density matrix $\left[\rho^\prime_{j'}\right]_{ab} \equiv         \left[\rho^\prime_{j'}\right]_{a_1a_2b_1b_2}$ on the ($5,4$) lattice has the form
\begin{equation}
    \begin{split}
        \left[\rho^\prime_{j'}\right]_{a_1a_2b_1b_2} &= \sum\limits_{\substack{c_1d_1\\e_1f_1}=1}^{q} \sum\limits_{\substack{c_2d_2\\e_2f_2}=1}^{n_j}
        {[\tilde{\cal C}_{j'}]}_{b_1b_2c_1c_2}
        {[\tilde{\cal C}_{j'}]}_{c_1c_2d_1d_2}\\
        &\times{[\tilde{\cal C}_{j'}]}_{d_1d_2e_1e_2}
        {[\tilde{\cal C}_{j'}]}_{e_1e_2f_1f_2}
        {[\tilde{\cal C}_{j'}]}_{f_1f_2a_1a_2},
    \end{split}
\end{equation}
where $a_1,b_1=1,2,\dots,q$, $a_2,b_2=1,2,\dots,n_j$, $a=n_j(a_1-1)+a_2$, and $b=n_j(b_1-1)+b_2$.
To prevent computational overflow in the numerical calculations, a maximal bond dimension $m$ restricts the growing spin degrees of freedom~\cite{DMRG1, DMRG2, DMRG3}. Hence, we restricted the exponential growth by setting $n_j = \min(q^j,m)$.

This restriction of the Hilbert-space dimension is carried out by selected eigenstates of $\rho_{j'}^{\prime}$. Having diagonalized the reduced density matrix
\begin{equation}
    \left[\rho^\prime_{j'}\right]_{ab} = \sum\limits_{i=1}^{qn_j} {[{\cal U}_{j+1}]}^{~}_{ai} {[\omega_{j'}]}_{i}^{~}\, {[{\cal U}_{j'}^{\dagger}]}_{ib}\,,
\end{equation}
we get eigenvalues ${[\omega_{j'}]}_{i}^{~}\geq0$ and their corresponding column eigenvectors in matrix ${\cal U}$ (let the eigenvalues be ordered decreasingly, i.e., ${[\omega_{j'}]}_{1}^{~}\geq{[\omega_{j'}]}_{2}^{~} \geq\dots\geq {[\omega_{j'}]}_{qn_j}^{~}$). By additional local normalization of $\tilde{\cal C}_{j'}$, we can prepare such $\rho^\prime_{j'}$ that satisfies ${\rm Tr}\, \rho^\prime_{j'} = \sum_{i=1}^{qn_j} {[\omega_{j'}]}_{i}^{~} =1$, where $\dim(\rho^\prime_{j'})=qn_j$. By construction, $\rho^\prime_{j'}$ is a real non-symmetric matrix for odd numbers of $p$ and has to be symmetrized: $\frac{1}{2} \left({\left[\rho^\prime_{j'}\right]}_{ab} + {\left[\rho^\prime_{j'}\right]}_{ba} \right) \to {\left[\rho^\prime_{j'}\right]}_{ab}$ to prevent from having the isometry in complex algebra.

To specify the renormalization isometry ${\cal U}_{j'}$, we keep such eigenvectors $\sum_{a=1}^{qn_j} {[{\cal U}_{j'}]}_{ai}|a\rangle$ that correspond to the $n_j$ largest eigenvalues of $\rho^\prime_{j'}$. Hence, the isometry
\begin{equation}
    {\cal U}_{j'} = \sum\limits_{a=1}^{q n_j} \sum\limits_{i=1}^{n_j} {[{\cal U}_{j'}]}_{a i}^{~} |a\rangle\langle i|
\end{equation}
fixes the exponentially growing Hilbert space to be bounded by $m$ after a couple of initial iterations.

Finally, we renormalize the extended tensors $\tilde{\cal C}_{j+1}$ and $\tilde{\cal P}_{j+1}$ by applying the isometry ${\cal U}_{j+1}$ (being a rectangular $q n_j \times n_j$ matrix of the $\rho_{j+1}^{\prime}$ eigenstates) as in Eq.~\eqref{renp4}.

\section{Spontaneous magnetization}
\label{SpMg}
\begin{figure}[tb]
{\centering\includegraphics[width=\linewidth]{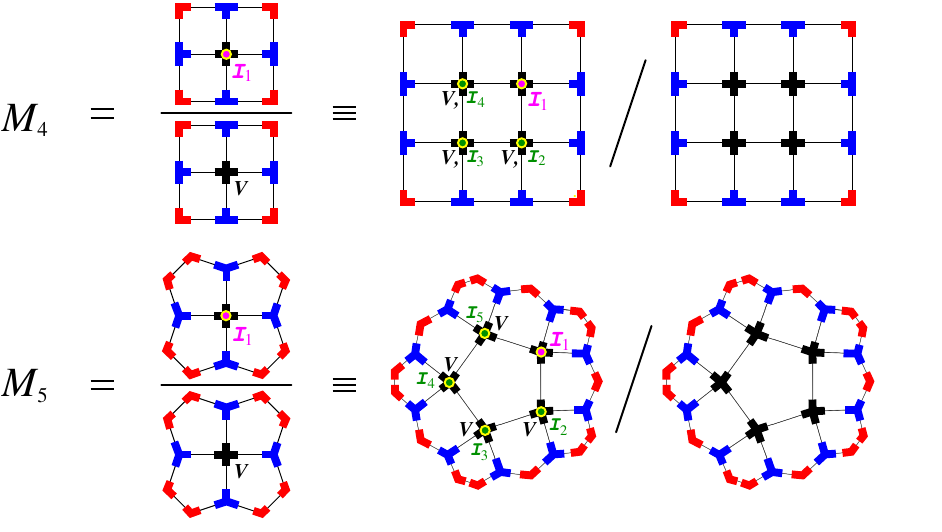}}
  \caption{Two examples of the spontaneous-magnetization calculation on the square $(4,4)$ and the pentagonal $(5,4)$ lattices as expressed in Eq.~\eqref{Mpq2}. The first part of the two expressions $M_4$ and $M_5$ uses a single impurity tensor ${\cal I}_1$ in the numerator to measure ${\cal M}_p$ at the central spin only. (The denominator is the normalizing partition function.) The first expression makes the CTMRG algorithm substantially faster than the second. Both lead to identical magnetizations in the thermodynamic limit.}
  \label{Mag_CTMRG}
\end{figure}

We can specify many ways to calculate the spontaneous magnetization ${\cal M}_p$. In Fig.~\ref{Mag_CTMRG}, we provide a graphical interpretation of obtaining two of them ${\cal M}_4$ (top) and ${\cal M}_5$ (bottom). The red and blue symbols refer to ${\cal C}_{k}$ and ${\cal P}_{k}$, respectively. The equivalent expression for the spontaneous magnetization (to that shown in Fig.~\ref{Mag_CTMRG}) for arbitrary ($p,4$) is
\begin{equation}
{\cal M}_p
=\frac{ {\rm Tr}\left[ {\cal I}_1\, {\cal P}_{k}^{4} {\cal C}_{k}^{4(p-3)} \right] }
      { {\rm Tr}\left[ {\cal V}\, {\cal P}_{k}^{4} {\cal C}_{k}^{4(p-3)} \right] }
      \equiv
      \frac{ {\rm Tr}\left[ {\cal I}_1\,{\cal V}^{p-1}_{~}\, {\cal P}_{k}^{2p} {\cal C}_{k}^{p(2p-7)} \right] }
      { {\rm Tr}\left[ {\cal V}^p_{~}\, {\cal P}_{k}^{2p} {\cal C}_{k}^{p(2p-7)} \right] },
\label{Mpq2}
\end{equation}
where the central spin ${\cal S}_{\sigma,\vartheta}$ has been absorbed in the impurity tensor defined in Eq.~\eqref{Iabcd} of the vertex structure of TN~\cite{HOTRG}. If focusing on the ($5,4$) lattice, the magnetization ${\cal M}_5$ expressed on the left has a lattice geometry where the lattice has a spin vertex ${\cal V}$ in its center while the lattice center forms a polygon (pentagon) on the right. Each one has strong and weak properties. Whereas the expression shown on the left-hand side is computationally more efficient and requires much less time to evaluate ${\cal M}_5$, the one on the right side follows the expansion scheme in Fig.~\ref{C1-C2} and can be used to calculate the nearest correlations, such as the internal energy, etc.

For the calculation of ${\cal M}_p$, we used the one on the left. To calculate the identical magnetization, the position of the impurity tensor ${\cal I}_\ell$ can be arbitrarily placed on the central polygon where $\ell=1,2,\dots,p$, as illustrated in Fig.~\ref{Mag_CTMRG} and Eq.~\eqref{Mpq2} on the right-hand sides. The remaining vertices occupy the polygon by the vertex-type tensors ${\cal V}$. To point it out again, the two approaches yield identical results for ${\cal M}_p$ as the number of iterations grows.

If increasing the number of impurity tensors ${\cal I}_{\ell}$, we can evaluate two-, three-, up to $p$-body correlation functions around the central polygon. For instance, the bulk internal energy per spin on the $(p,4)$ lattice is proportional to the nearest-neighbor (two-body) correlation function
\begin{equation}
    \langle {\cal S}_{\sigma_1\sigma_2} \rangle \propto -J \frac{ {\rm Tr}\left[ {\cal I}_1{\cal I}_2\,{\cal V}^{p-2}_{~}\, {\cal P}_{k}^{2p} {\cal C}_{k}^{p(2p-7)} \right]} 
    { {\rm Tr}\left[ {\cal V}^p_{~}\, {\cal P}_{k}^{2p} {\cal C}_{k}^{p(2p-7)} \right] }.
\end{equation}
The specific heat corresponds to the derivative of the internal energy $\langle {\cal S}_{\sigma_1\sigma_2} \rangle$ with respect to temperature $T$. The non-analyticity at the specific heat maximum is associated with the phase transition~\cite{weightpq}.

\section{Free-energy calculation}
\label{FreeEng}
As briefly sketched in Sec.~\ref{ThQu}, we now derive the free energy per spin by calculating the partition function and the number of spins at each iteration step $j$. The partition function diverges extremely fast with the increasing lattice size, especially, in the hyperbolic geometry where the number of spins grows exponentially with $j$.

Normalizing ${\cal P}_j$ and ${\cal C}_j$ at each iteration step $j$ is thus an inevitable condition. Following the ideas of the weight representation~\cite{weightpq}, we normalize the corner transfer tensors ${\bar{\cal C}}_j = {{\cal C}_j}/{\vert\vert{\cal C}_j\vert\vert}_{\max}$ and the transfer tensor ${\bar{\cal P}}_j = {{\cal P}_j}/{\vert\vert{\cal P}_j\vert\vert}_{\max}$ at each iteration step $j=1,2,3,\dots,k$. We use the maximum norm ${\vert\vert \cdot\vert \vert}_{\max}$ that searches all the tensor elements and finds the largest element in absolute value $x_j={\vert\vert{\cal C}_j\vert\vert}_{\max}$ and $y_j={\vert\vert{\cal P}_j\vert\vert}_{\max}$. The maximum norm is applied to the extended and renormalized tensors ${[{\cal C}_{j+1}]}_{a'b'}$ and ${[{\cal P}_{j+1}]}_{a'fb'}$ to obtain $x_{j+1}$ and $y_{y+1}$, respectively, cf. Fig.~\ref{renormalization}.

Consider the corner transfer tensor at the third iteration step ${\cal C}_{3}$ on the $(5,4)$ lattice. Using the recurrence relations in Eq.~\eqref{extp4short} for $p=5$, the corner transfer tensor at the third iteration step is recursively traced back to the initial tensors ${\cal C}_{1}$ and ${\cal P}_{1}$ to retrieve all the normalization values $x_1,x_2,x_3,y_1,y_2,$ and $y_3$. Then, we find
\begin{equation}
\begin{split}
    {\bar{\cal C}}_3^{~} & = \frac{{\cal C}_3^{~}}{x_3^{~}}
    = \frac{{\cal V} {\bar{\cal P}}_2^{2}{\bar{\cal C}}_2^{3}}{x_3^{~}}
     =\frac{{\cal V} {\cal P}_2^{2}{\cal C}_2^{3}}{y_2^2 x_2^{3}x_3^{~}} \\
    & = \frac{{\cal V}
    {\left({\cal V}{\bar{\cal P}}_1
    {\bar{\cal C}}_1^{~}\right)^2}
    {{\left({\cal V} {\bar{\cal P}}_1^{2}{\bar{\cal C}}_1^{3}\right)}^3}
    }{y_2^2 x_2^{3} x_3^{~}}
    = \frac{{\cal V}^6{\cal P}_1^{8}{\cal C}_1^{11}}
    {(y_1^{8} y_2^{2} y_3^{0}) (x_1^{11}x_2^{3}x_3^{1})}\,.
\end{split}
\label{C_norm_54}
\end{equation}
Let us recall that the initial tensors ${\cal V}$, ${\cal P}_{1}$, and ${\cal C}_{1}$ are associated with the single spin (see Fig.~\ref{C1-C2}). Therefore, the sum of their powers in the numerator of the bottom line in Eq.~\eqref{C_norm_54} yields $6+8+11=25$. That
corresponds to the corner tensor ${\bar{\cal C}}_{k=3}$ containing 25 spins. (We can also get the identical result with 25 spins if summing the powers of $x_j$ and $y_j$ in the denominator.)

Analogously, ${\bar{\cal C}}_{2} = {\cal V}^{1}{\cal P}_1^{2}{\cal C}_1^{3}/(y_1^2 y_2^0 x_1^3 x_2^1)$ contains $6$ spins which agrees with Fig.~\ref{C1-C2} enclosed by the dotted line. The number of spins at given step $k$ and lattice geometry $(p,4)$ has to be multiplied by $p$. Thus, the normalization factors and their powers are used to calculate the partition function and information in the powers can be used to get the number of spins.

Let $a_{k+1-j}$ and $b_{k+1-j}$ denote the powers of $x_j$ and $y_j$, respectively, recalling that $j=1,2,\dots,k$. For instance, the index ordering in Eq.~\eqref{C_norm_54} corresponds to the exponents $y_{1}^{b_{3}} y_{2}^{b_{2}} y_{3}^{b_{1}} x_{1}^{a_{3}} x_{2}^{a_{2}} x_{3}^{a_{1}}$. We thus obtained the second set of recurrence equations for the $(5,4)$ lattice
\begin{equation}
\begin{split}
a_{j+1} & =3a_{j}+b_{j}\, , \\
b_{j+1} & =2a_{j}+b_{j}\, .
\end{split}
\end{equation} 
After a longer analysis, we can further generalize the second set of the recurrence equations to get them for an infinite class of the lattices $(p\geq4,4)$
\begin{equation}
\begin{split}
a_{j+1} & =(2p-7)a_{j}+(p-4)b_{j}\, ,\\
b_{j+1} & =2a_{j}+b_{j}\, .
\label{ab}
\end{split}
\end{equation}
We initialize them $a_1=1$ and $b_1=0$ irrespective of $p$. These powers are necessary for evaluating the number of all spins on the $(p,4)$ lattice after $k$ iterations, i.e., we count the number of the vertices (tensors ${\cal V}$)
\begin{equation}
\label{Npk}
{\cal N}_{p,k} = p\sum\limits_{j=1}^{k} \left( a_j + b_j \right) \, .
\end{equation}
where the prefactor $p$ corresponds to dividing the lattice into identical $p$ parts (corners) ascribed to the corner transfer tensors ${\cal C}_{k}$, as shown in Fig.~\ref{extension}.

Equations~\eqref{extp4short}, \eqref{ab}, \eqref{Npk} are required for evaluating the free energy per spin
\begin{equation}
    {\cal F}_{p,k}=-\frac{k_{\rm B} T}{ {\cal N}_{p,k}}\ln {\cal Z}_{p,k}.
    \label{fe2}
\end{equation}
Following Eqs.~\eqref{extp4short}, \eqref{pf}, and \eqref{C_norm_54}, we can calculate $\ln {\cal Z}_{p,k}$ recursively by means of the norms and their associated powers $x_j^{a_{k-j+1}}$ and $y_j^{b_{k-j+1}}$
\begin{equation}
\begin{split}
    \ln {\cal Z}_{p,k} & = \ln \hspace{-0.2cm} \sum_{a_1 a_2 \cdots a_p} \hspace{-0.2cm} {[{\cal C}_{k}]}_{a_1 a_2} {[{\cal C}_{k}]}_{a_2 a_3} \cdots {[{\cal C}_{k}]}_{a_p a_1}\\
    & = \ln {\rm Tr}\, {\left({\cal C}_{k}^{p}\right)} = \ln {\rm Tr}\, {\left[ {\cal V} {\cal P}^2_{k-1} {\cal C}^{2p-7}_{k-1} \right]}^{p} = \cdots \\
    & = \ln {\rm Tr}\, {\left[ ({\cal V}^{n_v}_{~} {\cal P}^{n_p}_{1}  {\cal C}^{n_c}_{1})  \right]}^{p} \\
    & = \ln {\rm Tr} \left[ {\bar{\cal C}}_k^{~} \left( y_{1}^{b_k}y_{2}^{b_{k-1}}\hspace{-0.2cm} \dots y_{k}^{b_1}x_{1}^{a_k}x_{2}^{a_{k-1}}\hspace{-0.2cm} \dots x_{k}^{a_1}\right)\right]^p \\
    & =  \ln {\rm Tr} {(\bar{\cal C}_{k}^{p})} + p\sum_{j=1}^k {b_{k-j+1}} \ln (y_{j})+{a_{k-j+1}} \ln (x_{j}).
    \label{lnZ}
\end{split}
\end{equation}
Notice that $p(n_v+n_p+n_c) = {\cal N}_{p,k}$ from Eqs.~\eqref{C_norm_54} and \eqref{Npk}.
The first term in the bottom line in Eq.~\eqref{lnZ} is a small real number with upper bound $0 < \ln {\rm Tr}\, {(\bar{\cal C}_{k}^{p})} \ll 2p\ln(qm)$. If the first term in Eq.~\eqref{fe2} enters the free energy, the ratio ${\cal N}_{p,k}^{-1} \ln {\rm Tr}\, {(\bar{\cal C}_{k}^{p})} \to 0$ exponentially fast (the max norm applied to the tensor $\bar{\cal C}_{k}$).

\section{Free-energy asymptotics}
\label{FEas}

\subsection{Low-temperature limit}
\label{FEzero}

The uniform ferromagnetic ordering ($J=1$) at zero temperature leads to the $q$-fold degeneracy (unless the degeneracy is removed by spontaneous symmetry breaking in the thermodynamic limit). Equivalently, the symmetry can be broken by imposing small magnetic fields $h$ and $b$ (including numerical round-off errors). Then, one of the $q$ uniform orderings is chosen. The contribution to the $q$-state clock/Potts Hamiltonians enters as the number of bonds for the spin-spin interaction and the number of spins associated with the magnetic fields.

To evaluate the free energy as a function of the iteration step $k$, we compute the total number of spins ${\cal N}_{p,k}$, the number of spins on the boundary shell $n_{p,k}^{\rm BS}$, and the total number of bonds ${\cal B}_{p,k}$ as a function of the iteration step $k$. We can also ascribe $k$ to the number of shells with $k$ being the outermost spin shell and the first layer in the lattice center.

As derived in Appendix~\ref{FreeEng}, the construction of a $(p,4)$ lattice is described by the recursive relations in Eq.~\eqref{ab}. The total number of spins, derived in Eq.~\eqref{Npk}, is
\begin{equation}
    {\cal N}_{p,k} = p\sum\limits_{j=1}^{k} \left( a_j + b_j \right) = \underbrace{p(a_k + b_k)}_{n_{p,k}^{\rm BS}} + p\sum\limits_{j=1}^{k-1} \left( a_j + b_j \right),
\end{equation}
where $n_{p,k}^{\rm BS}$ is the number of spins on the outermost boundary shell (BS) at the final iteration step $k$. The total number of bonds, ${\cal B}_{p,k}^{~}$, is given by the number of bonds on all shells (which is equal to the number of spins on the shells ${\cal N}_{p,k}^{~}$) and the bonds connecting the shells. Between two shells corresponding to iteration steps $j$ and $j+1$, there is one bond per each tensor ${\cal P}_j$ and two bonds per each tensor ${\cal C}_j$. Thus, the total number of bonds at iteration step $k$ is:
\begin{equation} 
    {\cal B}_{p,k}^{~} = {\cal N}_{p,k}^{~} + p\sum\limits_{j=1}^{k-1} (2a_{j} + b_{j}) 
\end{equation}
On the square lattice, $(p=4)$, the recursive relations in Eq.~\eqref{ab} reduce to $a_{j+1}=a_{j}=1$ and $b_j=2j-2$. Hence
\begin{equation}
\begin{split}
    n_{4,k}^{\rm BS} &= 4(2k-1), \\
    {\cal N}_{4,k} &= 4\sum\limits_{j=1}^{k} \left( a_j + b_j \right) = 4\sum\limits_{j=1}^{k} \left( 2j-1 \right) = (2k)^2, \\
     {\cal B}_{4,k} &= 4\sum\limits_{j=1}^{k} \left( 2j-1 \right) + 4\sum\limits_{j=1}^{k-1} 2j = 4k(2k-1).
\end{split}
\label{4nNB}
\end{equation}

Now, we take the limit of the Bethe lattice ($p\gg 1$). The recurrence relations are $a_{j+1}=p(2a_j + b_j)=p\,b_{j+1}$. Then, $a_j=p\,b_j=2p(2p+1)^{j-2} \overset{ p\ggg1}{\longrightarrow} (2p)^{j-1}$ and we get
\begin{equation}
    \begin{split}
        n_{p,k}^{\rm BS} = \frac{1}{2}(2p)^k &,\\
        {\cal N}_{p,k}^{~} = \frac{1}{2}(2p)^k & + 
        \sum\limits_{j=1}^{k-1} \frac{c_j}{(2p)^{j}} \ \overset{ p\ggg1}{\longrightarrow}\ \frac{1}{2}(2p)^k, \\
        {\cal B}_{p,k}^{~} = \frac{1}{2}(2p)^k & + 
        \sum\limits_{j=1}^{k-1}\frac{d_j}{(2p)^{j}} \ \overset{ p\ggg1}{\longrightarrow}\ \frac{1}{2}(2p)^k,
    \end{split}
\label{nNB}
\end{equation}
where $c_j$ and $d_j$ are non-negative real numbers. With the results listed in Eqs.~\eqref{nNB} and \eqref{4nNB}, we evaluate the free energy per spin in the thermodynamic limit ($k\to\infty$) and at zero temperature $T\to0$. The spontaneously broken ferromagnetic spin alignment is uniformly ordered and Hamiltonian from Eq.~\eqref{H} reduces to
\begin{equation}
    {\cal H}_p[\sigma] = -J {\cal B}_{p,k}^{~} - h\, {\cal N}_{p,k}^{~} - b\, n_{p,k}^{\rm BS},
\end{equation}
where the interaction term $J$ acts on all the lattice bonds ${\cal B}_{4,k}^{~}$, the uniform magnetic field $h$ is applied to all spins ${\cal N}_{4,k}^{~}$, and the field $b$ is imposed on the boundary spins $n_{4,k}^{\rm BS}$ only. The free energy per spin for the square lattice ($4,4$) in Eq.~\eqref{fe} asymptotically ($T=0$) becomes
\begin{equation}
    \begin{split}
        {\cal F}_{4,\infty} & = \lim\limits_{k\to\infty} -\frac{k_{\rm B} T} {{\cal N}_{4,k}}\ln {\cal Z}_{4,k} \\
        & = \lim\limits_{k\to\infty} -\frac{k_{\rm B} T} {{\cal N}_{4,k}}\ln \sum\limits_{[\sigma]} \exp\left(-{{\cal H}_p[\sigma]}/{k_{\rm B} T} \right) \\
        & = \lim\limits_{k\to\infty} - \frac{J {\cal B}_{4,k}^{~} + |h|\, {\cal N}_{4,k}^{~} + |b|\, n_{4,k}^{\rm BS}} {{\cal N}_{4,k}} \\
        & = -2J - |h|.
    \end{split}
\end{equation}
The boundary field $b$ does not affect the free energy in the thermodynamic limit because $n_{4,k}^{\rm BS}$ grows linearly and $b$ is irrelevant on the Euclidean lattices.

Substantially different behavior occurs on the hyperbolic lattices. For the Bethe lattice $(\infty,4)$ we obtain
\begin{equation}
    \begin{split}
        {\cal F}_{\infty,\infty} & = \lim\limits_{\substack{k\to\infty\\ p\to\infty}} - \frac{k_{\rm B} T} {{\cal N}_{p,k}}\ln {\cal Z}_{p,k} \\
        & = \lim\limits_{\substack{k\to\infty\\ p\to\infty}} - \frac{J {\cal B}_{p,k}^{~} + |h|\, {\cal N}_{p,k}^{~} + |b|\, n_{p,k}^{\rm BS}} {{\cal N}_{p,k}} \\
        & = -J - |h| - |b|.
    \end{split}
\end{equation}

Finally, the asymptotic behavior of free energy at zero temperature is confirmed and agrees with the data shown in Fig.~\ref{FreeEnergy}. For the ferromagnetic interaction $J=1$ and field $h=0$, we get ${\cal F}_{4,\infty}=-2$ and ${\cal F}_{\infty,\infty} = -1 - |b|$.

\subsection{High-temperature limit}
\label{FEinf}

Thermodynamic entropy $S_{\rm th}$ is related to the temperature derivative of the free energy, see Eqs.~\eqref{fe} and \eqref{Sth},
\begin{equation}
    S_{\rm th} = \lim\limits_{T\to \infty}-\frac{\rm d} {{\rm d} T} \left(-\frac{k_{\rm B}T}{{\cal N}_{p,k}}\ln {\cal Z}_{p,k} \right).
\end{equation}
To evaluate the partition function, we realize that $\exp(-{\cal H}_p/k_{\rm B}T)\to1$ in the limit $T\to\infty$ irrespective of the type of Hamiltonian and lattice. Then the partition function reads as 
\begin{equation}
    {\cal Z}_{p,k} = \lim\limits_{T\to\infty}\underbrace{
    \sum_{\sigma^{~}_1} \sum_{\sigma^{~}_2}  \ \cdots \sum_{\sigma^{~}_{{\cal N}_{p,k}} }}_{ q^{{\cal N}_{p,k}} 
    \text{\ configs}}
    \exp\left(-\frac{{\cal H}_p[\sigma]}{k_{\rm B}T}\right) = q^{\,{\cal N}_{p,k}}.
\end{equation}
The partition function grows exponentially with the number of spins ${\cal N}_{p,k}$  for the $q$-state spin $\sigma_j = 0,1,\dots,q-1$. 
At high temperatures and $k_{\rm B}=1$, we obtain $\exp\left(S_{\rm th}\right) = q$, as plotted in the inset of Fig.~\ref{FreeEnergy}.



\end{document}